\documentclass[10pt,aps,prd,superscriptaddress,nofootinbib,nobibnotes,longbibliography,floatfix,twocolumn]{revtex4-2}

\usepackage{bm}
\usepackage{mathtools,
amsmath,
amssymb,
amsfonts,
mathrsfs,
chngcntr,
multirow}

\usepackage{float}
\usepackage{cancel}

\let\cc\corresponds
\let\corresponds\relax
\usepackage{mathabx}
\let\corresponds\cc

\usepackage[utf8]{inputenc}
\usepackage[T1]{fontenc}

\usepackage{soul}
\usepackage{tablefootnote}

\usepackage[dvipsnames]{xcolor}
\usepackage[unicode]{hyperref}
\hypersetup{colorlinks=true, citecolor=MidnightBlue,
            linkcolor=MidnightBlue, urlcolor=MidnightBlue, linktocpage=true}
\usepackage[normalem]{ulem}

\usepackage{graphicx}

\usepackage{enumerate}

\usepackage{comment}

\usepackage{accents}
\newlength{\dhatheight}

\begin{document}
\title{Binary black hole mergers in massive Einstein-scalar-Gauss-Bonnet gravity}

\author{Harry~L.~H.~Shum}
\email{likhangharry.shum@nottingham.ac.uk}
\affiliation{Nottingham Centre of Gravity,
Nottingham NG7 2RD, United Kingdom}
\affiliation{School of Mathematical Sciences, University of Nottingham,
University Park, Nottingham NG7 2RD, United Kingdom}

\author{Llibert Aresté Saló}
\email{llibert.areste@uib.cat}
\affiliation{Departament de F\'isica, Universitat de les Illes Balears, Palma de Mallorca, E-07122, Spain}
\affiliation{Institute of Applied Computing \& Community Code (IAC3), Universitat de les Illes Balears, Palma de Mallorca, E-07122, Spain}

\author{Miguel~Bezares}
\email{miguel.bezaresfigueroa@nottingham.ac.uk}
\affiliation{Nottingham Centre of Gravity,
Nottingham NG7 2RD, United Kingdom}
\affiliation{School of Mathematical Sciences, University of Nottingham,
University Park, Nottingham NG7 2RD, United Kingdom}

\author{Thomas~P.~Sotiriou}
\email{thomas.sotiriou@nottingham.ac.uk}
\affiliation{Nottingham Centre of Gravity,
Nottingham NG7 2RD, United Kingdom}
\affiliation{School of Mathematical Sciences, University of Nottingham,
University Park, Nottingham NG7 2RD, United Kingdom}
\affiliation{School of Physics and Astronomy, University of Nottingham,
University Park, Nottingham NG7 2RD, United Kingdom}

\begin{abstract}
We study fully non-linear binary black hole mergers in massive Einstein-scalar-Gauss-Bonnet (EsGB) gravity with a linear Gauss-Bonnet coupling function. In the absence of a well-developed initial data solver for EsGB gravity, we demonstrate that the conventional workaround solution for massless scalar fields results in confined junk radiation that contaminates late-time dynamics in the massive theory. We propose a solution to this problem by suitably modifying the equations of motion at the early stages of the simulation to tame such junk radiation, thereby allowing us to study the dynamics of the scalar field during the process of a black hole merger. This is followed by a preliminary analysis of the frequency content in the ringdown of the scalar and tensor modes. In particular, the quadrupole scalar mode exhibits a rich ringdown structure: while it is dominated by the spin-2 ``contamination'' frequency in the early stages, its late time morphology becomes primarily determined by the scalar mass frequency.
\end{abstract}

\maketitle

\section{Introduction}
The recent advent in gravitational wave (GW) astronomy~\cite{LIGOScientific:2016aoc, LIGOScientific:2017vwq} has opened up new possibilities in performing strong-field tests of General Relativity (GR). Thanks to future detectors, such as LISA~\cite{LISA:2017pwj}, the Einstein Telescope~\cite{Punturo:2010zz} and Cosmic Explorer~\cite{Reitze:2019iox}, we will soon have access to data from binary mergers that covers a wider region of parameter space, allowing for more comprehensive tests of GR and searches for new fundamental physics~\cite{Barausse:2020rsu,LISA:2022kgy,LISA:2024hlh,ET:2025xjr}.

In the modern viewpoint, extensions to GR can be systematically organised using the Effective Field Theory (EFT) framework, where modifications to low energy physics are captured by higher-derivative operators. One representative theory of this type is the Einstein-scalar-Gauss-Bonnet (EsGB) theory. This theory augments GR by introducing an extra scalar field and a non-minimal coupling between the scalar and the Gauss-Bonnet (GB) invariant. It belongs to the Horndeski class theories of gravity~\cite{Horndeski:1974wa}, which is the most general scalar-tensor action in four dimensions yielding second-order equations of motion (EoM). From a mathematical point of view, the fact that the EoMs of EsGB gravity remain second order despite the presence of a fourth derivative term in the theory (the GB invariant) removes the mathematical complications that come with analysing higher-order PDE systems. From a physics perspective, EsGB gravity can exhibit different phenomenology from that of GR in the strong gravity regime, while passing weak-field tests. For example, BHs in EsGB gravity can support a non-trivial scalar field profile around BHs (i.e. scalar hair)~\cite{Kanti:1995vq,Kanti:1997br, Sotiriou:2013qea,Sotiriou:2014pfa}, violating the no-hair theorem of BHs in GR. In certain ``flavours'' of EsGB gravity, BHs can also exhibit spontaneous scalarisation (or descalarisation) under favourable conditions~\cite{Doneva:2017bvd, Silva:2017uqg, Dima:2020yac, Herdeiro:2020wei, Doneva:2020nbb,Silva:2020omi,Pardoe:2026tcg}, allowing the existence of both GR-compatible BHs and hairy BHs depending on the BH mass/spin. These considerations motivate the study of EsGB gravity as an interesting alternative to GR.

Most studies of EsGB gravity consider a non-minimally coupled massless scalar field (we will refer to this as massless EsGB theory)~\cite{Corman:2022xqg,Corman:2024vlk,Corman:2025wun,Lara:2025kzj,Swain:2026dfr,AresteSalo:2026zzc,Capuano:2026lhs, Lara:2026xcs, Lara:2026dwh}.  The introduction of a scalar mass is a natural generalisation and can present additional phenomenology (see e.g.~\cite{Lam:2024wpq, Kuroda:2023zbz, Lam:2024azd, PhysRevD.108.064057} for studies in massive scalar-tensor theories). Firstly, the scalar mass contributes a new length scale to the theory. Interesting dynamics can arise when the length scale associated with the scalar mass (known as the Compton wavelength) becomes comparable to the characteristic length scales of strong gravity systems (e.g. the separation of BHs in a BBH merger). Secondly, the mass opens up the possibility of forming scalar bound states around BHs, potentially powered by a superradiant instability~\cite{PhysRevD.22.2323, Rosa:2009ei, Witek:2012tr, Hod:2012zza, Yoshino:2013ofa, ZOUROS1979139} (see~\cite{Brito:2015oca} for a review).\footnote{We will not study superradiance in this work due to the challenges it poses on our code's ability to remain stable throughout the long timescales required.} Thirdly, light massive scalar fields (e.g. axions) are also considered to be possible dark matter candidates (see e.g.~\cite{Niemeyer:2019aqm, 10.3389/fspas.2019.00047, Hui:2021tkt}), which motivated efforts to study its implications in the strong gravity regime~\cite{Zhang:2022rex, Choudhary:2020pxy, Bamber:2022pbs, Aurrekoetxea:2024cqd, Machet:2025vzt, Roy:2025qaa}.  Finally, from an observational perspective, a massive scalar provides a better chance of evading the strong observational constraints imposed on the massless theory~\cite{Sanger:2024axs, PhysRevD.106.124001} due to its natural mechanism for suppressing scalar radiation~\cite{Alsing:2011er, PhysRevD.93.064005, PhysRevD.108.064057}. The particular bounds for massive EsGB gravity have been obtained in~\cite{Xie:2024xex}.  These arguments provide a compelling theoretical motivation to study the predictions of massive EsGB gravity.

Stationary and spherically symmetric BH solutions in EsGB gravity with a massive scalar field have been constructed and studied~\cite{PhysRevD.99.104045,10.21468/SciPostPhysCore.7.4.069}, and recent efforts have also been made in modelling the dynamics of binary BHs (BBH) in massive EsGB gravity using the post-Newtonian framework~\cite{vanGemeren:2026yda}. In this paper, we extend the research in these lines by studying BBH mergers in massive EsGB gravity. Most of our study focuses on the case where the BHs have equal mass, in which the merger phase probes the highly non-linear spacetime dynamics. Solving the EoMs of the theory in this regime requires the use of Numerical Relativity (NR).

The application of NR techniques to modified theories of gravity has long challenged the research community due to several obstacles. First, the well-posedness properties of modified theories of gravity are in general not well-understood. It is therefore unclear whether suitable evolution formalisms can be found to perform stable long-time evolutions for such theories. This problem is partially alleviated by the recent breakthroughs in the development of the modified harmonic gauge (MHG)~\cite{Kovacs:2020pns, Kovacs:2020ywu} and the modified puncture gauge formalisms~\cite{AresteSalo:2022hua,AresteSalo:2023mmd, Shum:2025lgp}, which have provided well-posed formulations for a wide class of theories (including EsGB gravity) in the weakly coupled regime. The search for new evolution formulations of modified theories of gravity is still currently an active area of research~\cite{Figueras:2024bba, Figueras:2025wtx}.

The second obstacle to doing NR beyond GR comes from prescribing suitable initial data (ID) for evolution, as this requires solving the Hamiltonian and momentum constraints of a given theory. While methods for constructing initial data in GR are relatively well-established, developments in constructing modified gravity initial data are still at an early stage (see e.g.~\cite{Brady:2023dgu,Nee:2024bur} for recent work in this direction). To proceed with performing numerical studies of beyond-GR theories, a workaround solution is required. For example, in studies of massless EsGB theory, one usually assumes the scalar field and its time derivative to be initially vanishing. This reduces the EsGB constraint equations to the GR constraints, thereby allowing the use of GR initial data to initialise the simulation at the expense of generating transient junk scalar radiation as the scalar field settles to its quasi-stationary configuration during evolution. As we shall see, while this approach is successful for massless EsGB, it poses new problems for massive EsGB simulations due to the confinement of junk scalar radiation within the numerical domain. One of the main results of this paper is to propose an alternative workaround solution that involves dynamically controlling the spurious scalar radiation by suitably modifying the scalar EoM. Removing the confined junk radiation then allows us to consistently explore the dynamics of a BBH merger in massive EsGB gravity.

Since the removal of confined junk radiation results in a better estimate of the scalar field strength during a BBH merger, one can also study whether the scalar mass can leave an imprint in the scalar and gravitational waveforms from realistic BBH mergers of the massive EsGB theory. In particular, we perform a preliminary analysis of the ringdown frequency content in the scalar/tensor waveforms. By identifying the physical origins of the various frequencies extracted from the ringdown spectrum, one can not only understand the observability of the scalar mass through various waveforms but also identify any potential couplings between the scalar and gravitational modes. This allows for a better understanding of the observability of beyond-GR effects in realistic waveforms.

This paper is structured as follows. In Section \ref{sec:theory}, we first review the massive EsGB theory.  In Section \ref{sec:numerical_setup}, we discuss the numerical set-up for the simulations considered in this work. We then demonstrate that the conventional initial data prescription method fails for massive EsGB in Section \ref{sec:conv_ID}, thereby proposing an alternative solution. In Section \ref{sec:sims_with_damp_BH}, we test this alternative method in the case of an isolated non-spinning BH, before applying the same method to study BBH mergers in Section \ref{sec:sims_with_damp_BBH}. Finally, we perform a preliminary analysis of the frequency content in the ringdown of the various scalar/tensor waveforms in Section \ref{sec:ringdown}.

We adopt the conventions used in Wald’s book. Greek letters $\mu, \nu,...$ denote spacetime indices which run from 0 to d; and the $(l,m)$ modes of the scalar and gravitational waves (i.e. the $\Psi_4$ scalar) are denoted by $\phi_{lm}$ and $\Psi_{4,lm}$ respectively. We set $c=G=1$.

\section{The theory}\label{sec:theory}
The massive Einstein-scalar-Gauss-Bonnet action reads 
\begin{eqnarray}
    S&=\displaystyle\frac{1}{16\pi}\int d^4x \sqrt{-g} \left(R-\frac{1}{2}\partial_\nu\phi\,\partial^\nu \phi-\frac{1}{2}\mu^2 \phi ^2\right.\nonumber\\&\hspace{2cm}\left.+\lambda \, f(\phi)\,\mathcal{R}^2_{\rm GB}\frac{}{}\right)~,
\end{eqnarray}
where $\mu$ is the scalar mass, $\mathcal{R}^2_{\rm GB}=R^2-4R^{\alpha \nu}R_{\alpha \nu} + R^{\alpha \nu \rho \sigma}R_{\alpha \nu \rho \sigma}$ is the Gauss-Bonnet invariant, $f(\phi)$ is the coupling function of the theory, and $\lambda$ is the coupling constant between the scalar field and the Gauss-Bonnet invariant. A simple dimensional analysis argument reveals that $\lambda$ and $\mu$ have dimensions (length)$^2$ and (length)$^{-1}$ respectively.

In general, different choices of coupling functions $f(\phi)$ predict BHs with different properties~\cite{Doneva:2022ewd}. These functions can be broadly divided into two classes: if $f'(\phi)\neq0$, the scalar EoM implies that a trivial stationary solution of the scalar field cannot exist in the presence of non-zero curvature; hence, the BHs predicted by such coupling functions will always have scalar hair (implying GR BHs are not solutions of these theories)~\cite{Kanti:1995vq,Yunes:2011we,Sotiriou:2013qea,Sotiriou:2014pfa,Antoniou:2017acq}. On the other hand, if $f'(\phi_0)=0,$ for any given constant $\phi_0$, the theory admits both GR and scalarised BH solutions~\cite{Silva:2017uqg}. BHs in these theories exhibit interesting phenomenology such as scalarisation, where the BH migrates from a GR solution to a scalarised BH. For our study, we consider a linear coupling function $f(\phi)=\phi$. The linear coupling is naturally motivated from an EFT perspective as it describes the leading order of a non-minimal interaction between the scalar and spacetime geometry. Furthermore, as is evident from the scalar EoM,
\begin{equation}
    \nabla_{\nu}\nabla^{\nu} \phi -\mu^2 \phi=-\lambda \mathcal{R}^2_{\rm GB}~,
\end{equation}
BHs in this theory necessarily support a non-trivial scalar field profile, which allows the scalar field to play a more dynamical role in non-linear evolutions.

\section{Set-up}\label{sec:numerical_setup}
We study BHs in massive EsGB theory in two physical scenarios: an isolated non-spinning BH and a BBH. While most of our study concerns equal-mass BBHs, we will study one unequal mass BBH with ratio 1:2.

The simulations are performed using the modified CCZ4 (mCCZ4) formalism (see~\cite{AresteSalo:2022hua,AresteSalo:2023mmd} for further details) implemented in the public code \texttt{GRFolres}~\cite{AresteSalo:2023hcp}, which is an extension of \texttt{GRChombo}~\cite{Clough:2015sqa, Andrade:2021rbd} adapted for modified theories of gravity. The grid settings used for an isolated non-spinning BH and a BBH system are summarised, respectively, in Table~\ref{table:numerical_settings}. Note that the dimensionful physical quantities presented are measured in an arbitrary code unit $M$.
\renewcommand{\thempfootnote}{\arabic{mpfootnote}}
\begin{table}[h!]
\addtocounter{footnote}{1}
\xdef\mytablefn{\arabic{footnote}}
    \centering
    \begin{tabular}{|l|c|c|c|}
        \hline
        \textbf{System} & $dx_\text{finest}/M$ & $L/M$ & AMR levels \\
        \hline
        Non-spinning BH     & 1/32          & 512          & 8                   \\
        \hline
        BBH  & 1/96           & 1024          & 10                   \\
        \hline
\end{tabular}
\caption{\emph{Numerical set-up.} Numerical settings for an isolated non-spinning BH and a BBH system, respectively. The value of the spatial discretisation step $dx$ at the finest level is denoted as $dx_\text{finest}$, the length of the numerical domain is denoted as $L$, and the final column denotes the total number of adaptive mesh refinement (AMR) levels (including the coarsest level)\protect\hyperlink{table:numerical_settings}{\textsuperscript{\mytablefn}}. Note that our choice of resolution ensures that the temporal oscillations of fields can be sufficiently resolved~\cite{Traykova:2021dua}, i.e. satisfying the condition $T = 2\pi/\mu > n\, dt_{\text{coarsest}}$ with $n\gg 1$, where $n$ represents the number of time steps per period of oscillation due to the scalar mass. For instance, the smallest $n$ used in our simulations comes from the simulations of equal-mass BBH with $\mu M=0.2$, which takes the value of $n\sim23$.}
\label{table:numerical_settings}
\end{table}
\footnotetext[\mytablefn]{\hypertarget{table:numerical_settings}{}For the BBH simulations, we have taken the size of the finest level around each one of the punctures to be 1.5 times its mass. While sufficient for our purposes, a larger value would need to be chosen should one aim to do a more accurate study of the inspiral phase, which is needed for example to quantify the amount of dephasing with respect to GR~\cite{Corman:2025wun}.}

For the non-spinning BH simulations, we set the CCZ4 dissipation coefficients to $\kappa_1=1.0/M, \kappa_2=0, \kappa_3=1.0/M$ while for the BBH simulations we use $\kappa_2=-0.15$ instead. All physical fields are enforced with Sommerfeld boundary conditions except for the scalar field and its conjugate momentum, which are assumed to decay as a Yukawa potential $\sim e^{-\mu r}/(r/M)$ at the box boundary.

The use of the modified gauge in the mCCZ4 formalism introduces two additional parameters to be specified~\cite{Kovacs:2020pns,Kovacs:2020ywu}, which relate to the propagation velocities of the ``pure gauge'' and ``gauge-condition violating'' modes in the theory.  If we express the auxiliary metrics as
\begin{eqnarray}
    \tilde{g}^{\sigma \nu}&=g^{\sigma \nu} -b(x)n^\sigma n^\nu~,\\
    \hat{g}^{\sigma \nu}&=g^{\sigma \nu} -a(x)n^\sigma n^\nu~,
\end{eqnarray}
where $n^{\mu}$ is the normal covector to a given hypersurface; then the propagation velocities will be determined by the functions $a(x),b(x)$. For simplicity, we will treat them as constants with values set to $(a,b)=(0.2,0.4)$, as in the literature~\cite{AresteSalo:2023mmd}.

\begin{table}
    \centering
    \begin{tabular}{|c|c|c|c|c|c|}
        \hline
        \textbf{System} & $m_1/M$ & $m_2/M$ & $d/M$ & $p_1^i/M$ ($=-p_2^i/M$) \\
        \hline
        i    & 0.5          & 0.5          & 12.21358         & (-0.000510846, 0.0841746, 0) \\
        \hline
        ii  & $1$           & $0.5$          & 15      & (0, 0.128488732, 0)       \\
        \hline
\end{tabular}
\caption{\emph{BBH initial data.} 
We present the initial data for the BH masses ($m_1,m_2$), coordinate separation ($d$), and momenta ($p_1^i,p_2^i$) considered in our simulations, where system (i) refers to an equal-mass BBH and (ii) an unequal mass BBH. We note that the exact momenta used for most of our BBH simulations with massive scalars will differ from the presented values in the table due to the use of eccentricity reduction, which will be explained in Section \ref{sec:sims_with_damp_BBH}.
}
\label{table:ID}
\end{table}

Regarding the physical parameters, we set the coupling constant of our simulations to $\lambda=0.1 m_0^2$, where $m_0$ is the scale defined by the smallest BH in the system (e.g. in an equal-mass BBH with a total mass of $1M$, $m_0=0.5M$). As for the initial data, the isolated BH simulations are taken to have a BH mass of $1M$ and no spin, while the BBH simulations adopt Bowen-York initial data~\cite{Bowen:1980yu} generated by the standalone \texttt{TwoPunctures} code~\cite{Ansorg:2004ds}. The details regarding initial BH masses, coordinate separation and momenta are specified in Table~\ref{table:ID}. All BHs are taken to be initially non-spinning. For all cases, the scalar field and its momentum conjugate are set to be initially vanishing, such that the EsGB constraint equations reduce to those of GR. This allows us to use conventional GR initial data as the constraint satisfying data for our initial time slice. 

In the absence of a well-developed initial data solver for EsGB gravity (see~\cite{Brady:2023dgu,Nee:2024bur} for recent progress), this method of preparing initial data proves to be sufficiently satisfactory for BBHs in massless EsGB theory (up to the introduction of orbital eccentricity~\cite{Corman:2025wun}): the scalar field reacts to the curvature of the BHs and develops a quasi-stationary configuration within a transient timescale much shorter than orbital timescales, and the junk radiation generated in the process propagates away from the BH system before the system reaches merger phase. However, despite its success in the massless case, this method is not expected to work for a massive field.  Due to the confining property of the massive scalar potential, one can imagine that part of the junk radiation becomes trapped and does not propagate towards infinity. As we will show, this is indeed what happens, making the late time part of numerical simulations sensitive to the unphysical initial data used. This limits one's ability to extract reliable physics from a simulation using GR initial data, which motivates us to search for ways to tackle this problem. Our proposed method for controlling scalar waves will be described in Section \ref{sec:conv_ID}.

\section{Results}
We now present the main results of our study. We first discuss simulations of an isolated non-spinning BH, initiated with conventional initial data, to highlight the challenges in its use in simulations involving massive scalars. We then propose a workaround solution to this problem, which is tested using simulations of an isolated non-spinning BH before being applied in our study of BBH mergers.

\subsection{From Conventional Initial Data to Controlled Scalar Junk Radiation}\label{sec:conv_ID}
To illustrate how the conventional method of preparing scalar initial data would lead to confined junk radiation in the massive case, we first consider simulations of an isolated non-spinning BH that is initialised using the conventional method, i.e. setting the scalar field and its conjugate momentum $\Pi=n^{\nu} \nabla_{\nu} \phi$ to be initially vanishing (i.e., $\phi=\Pi=0$). The scalar is then sourced by the GB invariant, $\mathcal{R}^2_{\rm GB}$, at the start of the simulation, which in the massless theory will settle down to the stationary configuration within a short time scale after a transient phase (see e.g. Fig.~1 of \cite{Shum:2025lgp}).

\begin{figure}
    \centering
    \includegraphics[width=1\linewidth]{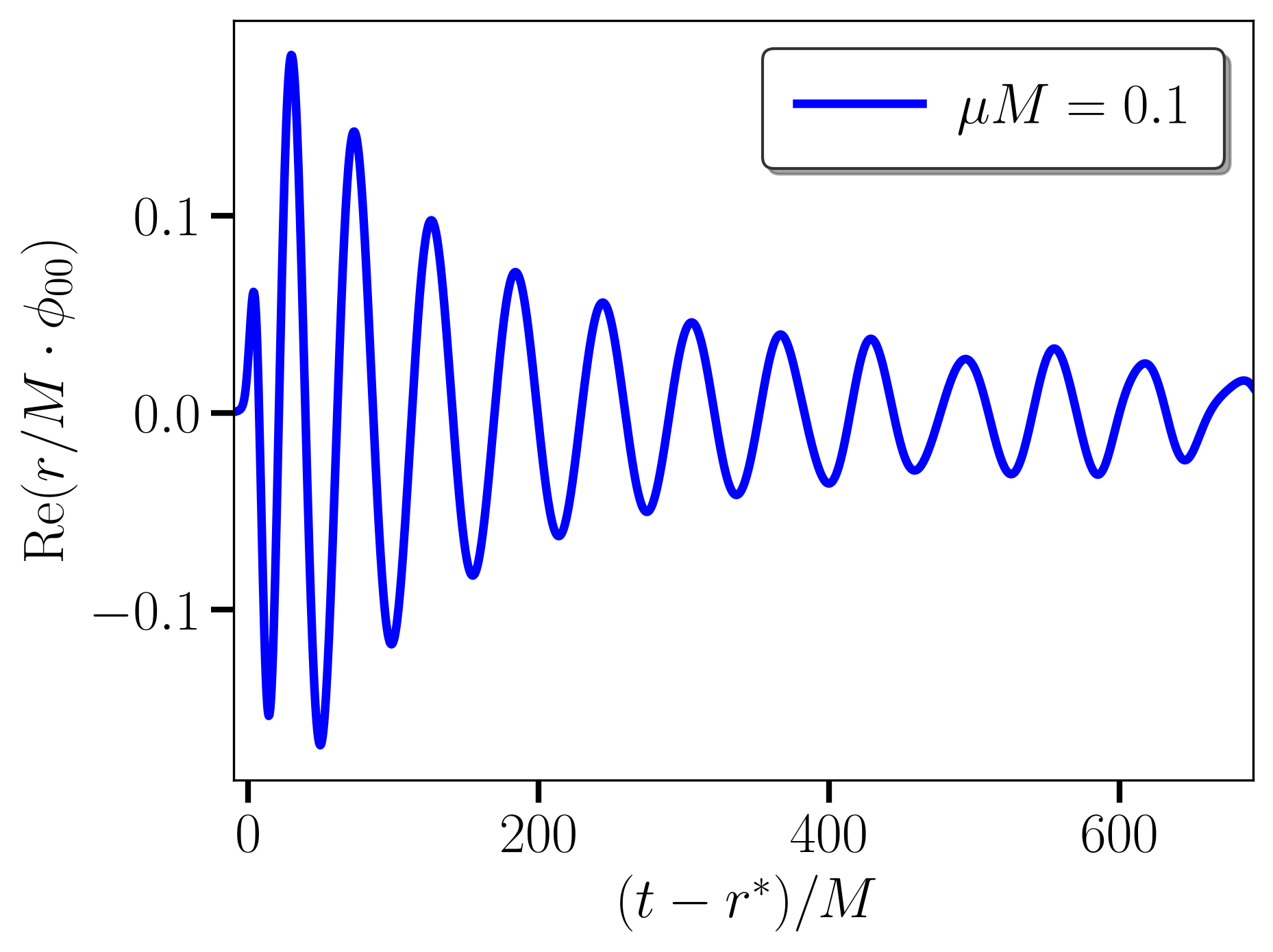}
    \caption{The (0,0) mode scalar waves with $\mu M=0.1$ extracted at $r=100M$ for a non-spinning BH. Instead of reaching a stationary configuration after some transient dynamics, the field keeps oscillating at a frequency determined by the scalar mass.}
    \label{fig:Kerr_nodamp_SW00}
\end{figure}

Due to the symmetry of our set-up, the dominantly excited scalar mode is the spherically symmetric $(0,0)$ mode, which we show in Fig.~\ref{fig:Kerr_nodamp_SW00}. Unlike the case of a massless scalar, the massive scalar never settles down to a stationary state within reasonable simulation time. Instead, the oscillations in the field persist up to late times in the simulation. Such oscillations are artefacts of using unphysical initial conditions for the scalar field: for EsGB gravity with linear coupling, the configuration $\phi=\Pi=0$ is not a stationary solution of the scalar EoM. Instead, if we were to initialise our simulations with a constraint-satisfying scalar field profile that corresponds to the stationary configuration, there would not be such significant oscillatory dynamics in the scalar field.

Performing a simple Fast Fourier Transform (FFT) with the scalar (0,0) mode reveals that the frequency content of the wave is dominated by the frequency $f\approx0.0163M^{-1}$, whose relation to the scalar mass can be identified when the number is converted to angular frequency $\omega=2\pi f\approx0.102M^{-1}\approx \mu$ (i.e. the inverse Compton wavelength of the scalar in geometric units up to a factor of $2\pi$)). This highlights the relation between the unphysical scalar waves and the scalar mass of the theory. The fact that the scalar field does not settle down to a stationary configuration implies that the simulations will be sensitive to the unphysical choice of initial data even at sufficiently late times, thus limiting ones ability to extract meaningful physics from the simulations \footnote{Despite the scalar field's problematic behaviour with the use of conventional ID, we find that certain geometric invariants might still display qualitative agreement with studies of stationary BHs in massive EsGB gravity (see Appendix \ref{sec:Kretsch} for a more detailed discussion on this aspect).}.

Following the previous discussion, it is clear that strategies to suppress the scalar junk radiation are necessary to prevent it from contaminating the late-time dynamics of the simulation.
This is especially relevant for highly dynamical systems such as a BBH merger, as the unphysical scalar dynamics can potentially impact the evolution of a BBH and limit the physical conclusions we can draw from the simulations. For instance, results from the previous section indicate that junk radiation contributes to the persistent unphysical dynamics in the (0,0) scalar mode radiation. If a similar feature is found in BBH simulations, one might be led to conclude that the (0,0) scalar mode is a viable energy emission channel during the inspiral phase for a massive scalar, when such radiation is only the consequence of using poor initial data. Moreover, the unphysical amplitudes of the scalar field triggered by junk dynamics can also lead to an overestimate of the ``amount'' of scalar at any given time. This could have implications on the evolution of the system as the scalar field backreacts onto the metric. In essence, analysing the data from these simulations might lead to false conclusions due to contamination from unphysical initial data.

To address this issue, we propose a new method to tame the trapped junk radiation. This approach modifies the scalar-field EoM so that we can dynamically connect the initial data configuration $\phi=\Pi=0$ to the quasi-stationary configuration admitted by the theory. In practice, this can be achieved by adding a damping term\footnote{Note that similar dissipative approaches are used in studies of viscous hydrodynamics and modified gravity to control high-frequency modes \cite{ISRAEL1976310,Cayuso:2017iqc,Besharat:2026rba}.} proportional to a time-dependent function $\kappa(t)$ to the scalar EoM, which yields
\begin{equation}
    \nabla_{\nu}\nabla^{\nu} \phi -\mu^2 \phi-\kappa(t) \Pi=-\lambda \mathcal{R}^2_{\rm GB}\,.
\end{equation}
Note that the explicit inclusion of $\Pi=n^{\mu}\nabla_{\mu}\phi$ in the scalar EoM introduces a preferred timelike direction, which is necessary for the dissipation of junk dynamics in the scalar.  We find that the damping term leads to the damping of the conjugate momentum. Moreover, it does not affect the hyperbolicity of the system, as it does not modify the EoM's highest-order derivative structure. As we evolve the system, we gradually switch off the damping factor by setting $\kappa(t) \rightarrow 0$. The scalar EoM thus relaxes to the original EoM, allowing us to smoothly evolve the constraint-satisfying ID towards the quasi-stationary scalar profile. Once the junk dynamics are sufficiently suppressed, we remove the damping term from the evolution equations such that the remaining evolution will be done using the original EoMs of EsGB gravity. In our current work, the damping term is removed at $t=500M$ for simulations involving a single BH; whereas for a BBH, a switch-off time at $t=400M$ and $800M$ are used for the scalar masses $\mu M=0.1, 0.2$, respectively. Note that the choice of switch-off time is obtained empirically by checking when the scalar dynamics is sufficiently damped, and is in general dependent on the scalar mass. As we shall see, this approach dissipates the junk radiation scalar effectively at the early stages of the simulations and confirms our expectation that using a more accurate initial configuration for the scalar field can remove certain unphysical dynamics originating from initial data artefacts.

As a first attempt in testing the effectiveness of the damping mechanism, we use the following prescription for the damping function $\kappa(t)$,
\begin{equation}
    \kappa(t)=\kappa_0 \left(1-\tanh{\frac{t}{T_0}}\right)\,,
\end{equation}
which has the property of smoothly decaying to zero over a timescale $T_0$. For the simulations we perform, this timescale is taken to be $T_0=150M$. This value is chosen such that it is longer than the characteristic timescale associated with the scalar mass, but at the same time allows the damping term to be switched off smoothly within reasonable simulation time. Choosing an optimal value for the damping constant $\kappa_0$ requires similar considerations: a value that is too small will render the damping effectively useless, while a damping that is too strong might over-suppress the growth of the scalar from the initial data $\phi=0$, as well as making the evolution equations stiff. Since our objective is to damp away the scalar junk radiation (which we have found to oscillate at the scalar mass frequency), we set $\kappa_0$ to be equal to the scalar mass $\mu$ (or values close to $\mu$), such that the timescale associated with the damping is comparable to that of the scalar waves.

\subsection{Numerical Simulations with Damping: From Isolated to Binary Black Holes}\label{sec:sims_with_damp}
In the simulations presented here, we will limit ourselves to a small sample of scalar masses to test the effectiveness of the damping method. Most of our analysis will be based on simulations with scalar mass $\mu M=0.1$. This choice is motivated by considering its implications on the BBH simulations: studies have shown that the scalar field can effectively affect orbital evolution when the orbital separation $d$ becomes comparable to the Compton wavelength $\sim 2\pi/\mu$~\cite{Aurrekoetxea:2023jwk, Machet:2025vzt}. In order to allow for the efficient development of scalar dynamics, we choose values of scalar masses such that the condition $2\pi/\mu \gtrsim d$ is satisfied. For our initial conditions of $d \sim 12M$, this criterion is satisfied by our choice of scalar mass $\mu M=0.1$. To verify that our conclusions are not specific to the choice of scalar mass or damping parameters, we will also consider BBH mergers with a scalar mass of $\mu M=0.2$. These choices of scalar masses translate to the physical masses $m_{\rm phy}=\hbar \mu/c$ of $1.34 \cdot10^{-11} (M_\odot/ M)\, \rm eV$ and $2.68 \cdot10^{-11}(M_\odot/M)\, \rm eV$ respectively.  Using the bounds obtained from observation runs O1-O3 of the LIGO-Virgo-KAGRA detectors~\cite{Xie:2024xex}, we find that for an equal-mass BBH with a total mass of $\sim 20M_\odot$ our choice of coupling strength ($\lambda/M^2=0.025$) and scalar mass ($\mu M=0.1$) translates to the physical values of $\sqrt{\alpha_{\rm GB}} \sim 1.75 \rm km$ \footnote{We note that the coupling constant ${\alpha_{\rm GB}}$ used in~\cite{Xie:2024xex} is related to our coupling constant $\lambda$ by $\lambda=\sqrt{16\pi}\alpha_{\rm GB}$.} and $m_{\rm phy}c^2 \sim6.70 \cdot10^{-13} \rm eV$, which lies in the regime that is not constrained by observations. In general, we expect that our choice of theory parameters will result in a system that is unconstrained by observations as long as the total mass satisfies $M\lesssim 20M_\odot$\footnote{Similar bounds for the unequal mass BBH with $\mu M=0.1$ and the equal-mass BBH with $\mu M=0.2$ yield $30M_{\odot}$ and $33M_{\odot}$, respectively.}. Note that this bound is a consequence of our particular choice of coupling strength and scalar mass, and larger bounds may be obtained by considering weaker couplings or larger scalar masses.

\subsubsection{Isolated non-spinning BH}\label{sec:sims_with_damp_BH}
Using the above damping method to control the junk scalar waves, we perform simulations of an isolated non-spinning BH to verify that the scalar-damped simulations produce expected physics, and the scalar field is not overly dissipated by our methods.

The most directly verifiable diagnostic that the scalar field assumes its stationary configuration is by checking its asymptotic spatial decay rate. As noted in the literature~\cite{PhysRevD.99.104045,10.21468/SciPostPhysCore.7.4.069} and as it can be easily verified, a massive scalar field decays at a rate of $\phi \sim e^{-\mu r}/(r/M)$ asymptotically far away from the BH. If the scalar-damped simulations were to represent accurate physics, this asymptotic behaviour in the scalar field should be reproduced. To verify this, one can plot the quantity $\ln(r/M\cdot|\phi|)$ as a function of $r$ and see whether the numerical value of the scalar mass can be recovered from its slope to a desirable accuracy. From our simulations, we take $r$ to be the coordinate distance from the centre of the BH puncture on a given spatial hypersurface, and $\phi$ the value of the scalar field extracted along $r$ at late times ($t=800M$). We note that the extraction line can be chosen to be in any arbitrary direction within a hypersurface due to the spherical symmetry of the problem. For simplicity, we have chosen it to align with one of the coordinate axes. Furthermore, by extracting the scalar profile at $t=800M$, we allow for sufficient time for any potential transient behaviour in the scalar field to settle down after the damping is switched off at $t=500M$. In order to ensure sufficient suppression of the junk scalar dynamics to allow for accurate fitting, we use a slightly stronger damping constant (relative to the scalar mass) of $\kappa_0=0.15/M$.

\begin{figure}
    \centering
    \includegraphics[width=\linewidth]{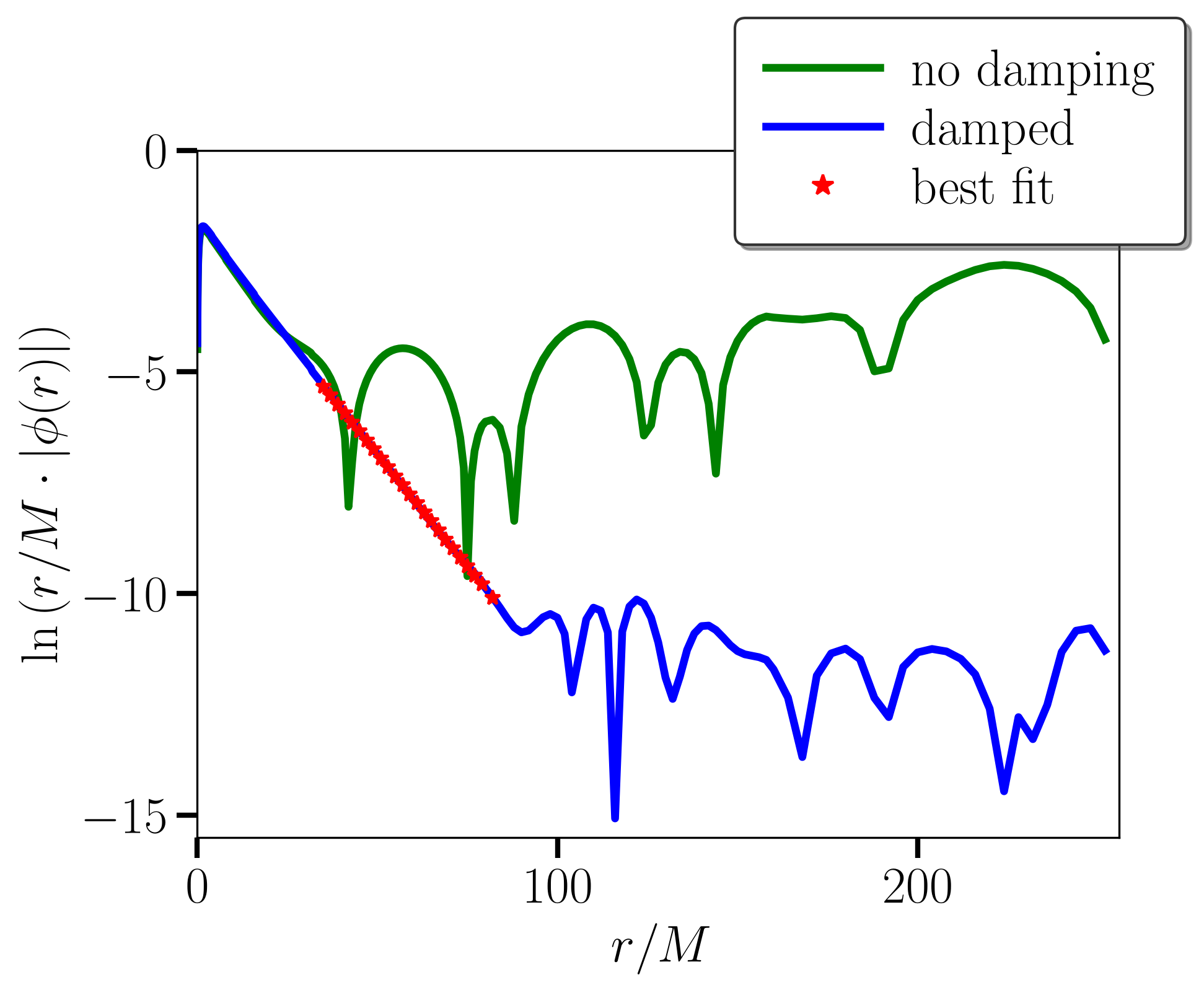}
    \caption{The radial profile of $\phi$ at $t=800M$, comparing the case where damping is used against no damping. To emphasise the asymptotic exponential decay behaviour of the damped-scalar field, we plot the quantity $\ln(r/M\cdot|\phi|)$ as a function of the coordinate radial distance $r$ from the BH. Due to the spherical symmetry in our spacetime, we can obtain this quantity by extracting the value of the scalar field along a radial line from the BH. Farther away from the BH, the scalar field demonstrates an exponential decay, up to a radius where the residual noise dominates. Performing a linear fitting between $r/M \in[35,85]$ produces a best-fit line (red) with a slope of -0.102, in strong agreement with the scalar mass $\mu M=0.1$.}
        \label{fig:phi_decay}
\end{figure}

As shown in Fig.~\ref{fig:phi_decay}, the quantity $\ln(r/M\cdot|\phi|)$ demonstrates a linear decay as a function of radius $r$, indicating that the scalar field is decaying as $e^{-\mu r}/(r/M)$ at large distances from the BH. Incidentally, for this choice of scalar mass, we find that one does not have to extract the field very far away from the BH to observe the expected decay rate. At distances $r\gtrsim 100M$, we find residual noise in the scalar field that is not dissipated by the damping. In principle, we believe they can be further controlled by switching off the damping over a longer time scale at the expense of higher computational costs. However, the small order of magnitude of these oscillations suggests that we can safely ignore them for practical purposes.
To verify that the slope of the linearly decaying segment in Fig.~\ref{fig:phi_decay} is given by the scalar mass, a simple linear fitting is performed. Within the fitting range of $r/M \in[35,85]$, we find that the best fit line produces a slope of $-0.102$, agreeing with the value of the scalar mass to a high precision. This suggests that our damping mechanism did not over-dissipate the scalar, and the scalar configuration we obtain after the damping procedure is indeed close to the expected stationary configuration.

\subsubsection{Binary BH}\label{sec:sims_with_damp_BBH}
Having tested the damping mechanism with an isolated non-spinning BH, we now apply it to study the dynamics of an equal-mass BBH. For this set of simulations, the damping constant will be chosen to be equal to the scalar mass, i.e. $\kappa_0=\mu$. We focus on two choices of scalar masses, $\mu M=0.1$ and 0.2, in addition to the massless EsGB and GR simulations for comparison.

\begin{figure}[h!]
    \centering
    \includegraphics[width=1\linewidth]{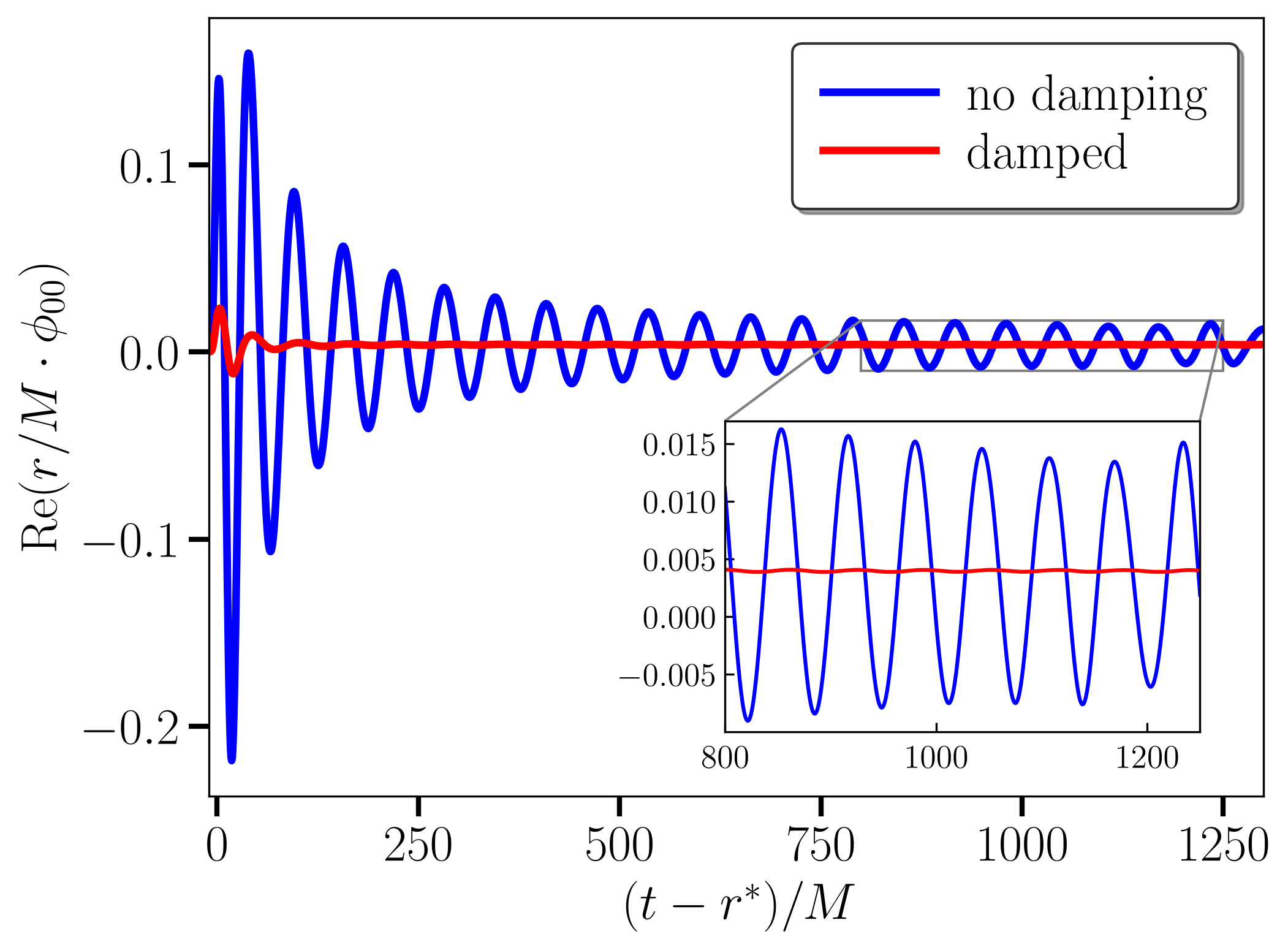}
    \caption{$\phi_{00}$ scalar waves of mass $\mu M=0.1$ extracted at $r=50M$ during the inspiral, comparing the case where no damping is used against the scalar-damped case with $\kappa_0=0.1 M^{-1}$. The inset shows the finer difference between the two ways of preparing initial data. While the conventional ID leads to significant confined radiation that is capable of switching the sign of the scalar field, the scalar-damped simulation shows minimal perturbations about a non-zero constant value $\approx 0.004$, which is of the same order of magnitude as $e^{-\mu r} \approx 0.007$ at an extraction radius of $r=50M$.}
    \label{fig:SW00_damped_vs_undamped}
\end{figure}

We first demonstrate that the damping successfully controls the junk scalar waves in BBH spacetimes, by considering simulations with scalar mass $\mu M=0.1$. Similarly to the non-spinning BH, the (0,0) scalar mode is the dominant scalar wave channel excited by the junk radiation. As shown in Fig.~\ref{fig:SW00_damped_vs_undamped}, while the undamped simulation results in trapped scalar radiation persisting up to late times, the damped simulation exhibits suppressed junk radiation already at early times during the inspiral phase. Moreover, the scalar field does not appear to be over-suppressed by the damping: if that were the case, one would expect the junk radiation to reappear once we remove the damping term, as the scalar will reconfigure itself in an attempt to reach the quasi-stationary configuration (similar to how the scalar field develops from trivial initial data to a non-trivial configuration in the undamped simulations). The fact that the scalar wave amplitude remains approximately constant after the damping term is removed at $t=400M$ suggests that we are sufficiently close to the quasi-stationary configuration towards the end of the damping procedure.

\begin{figure}
    \centering
    \includegraphics[width=1\linewidth]{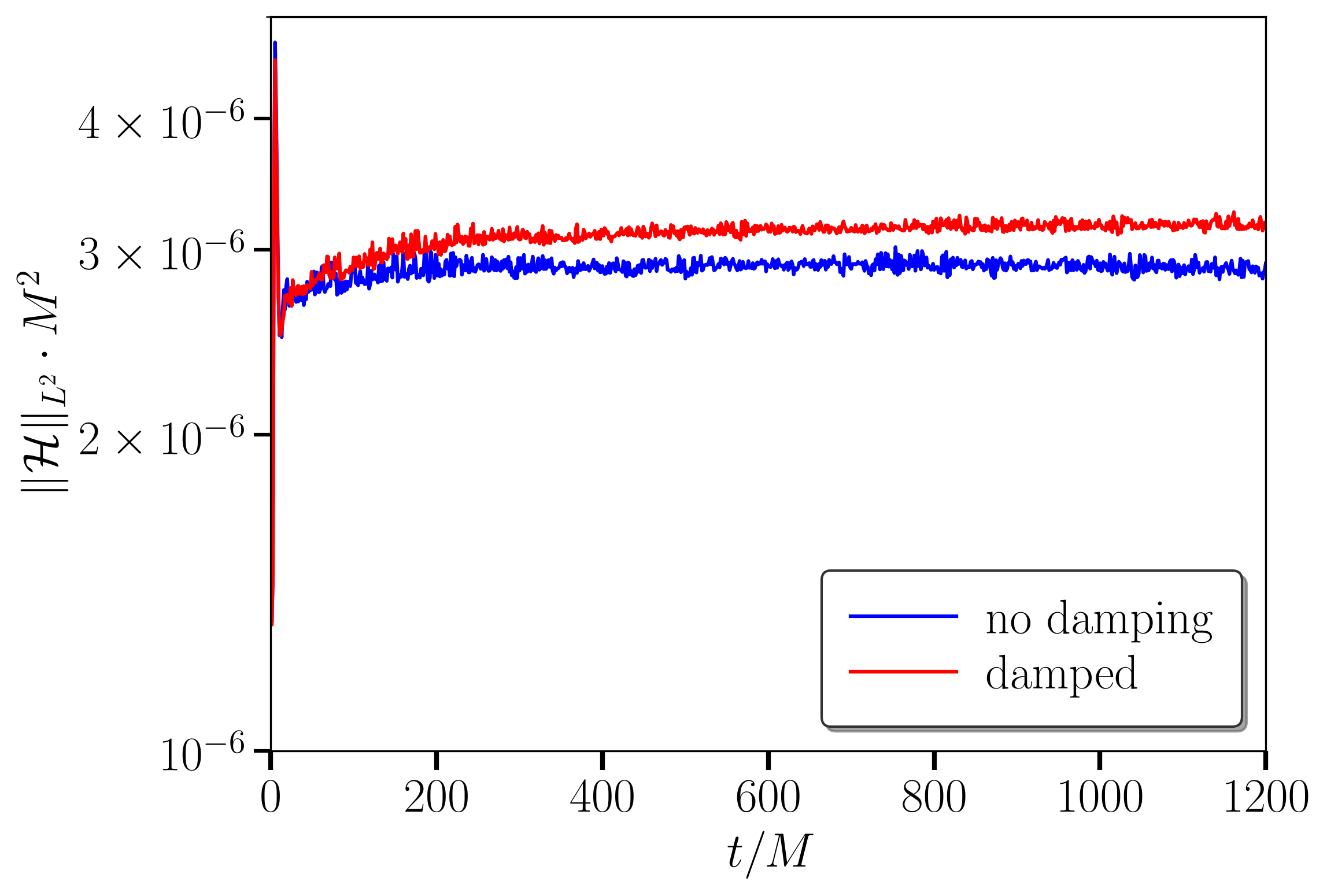}
    \caption{The evolution of the volume-normalised $L_2$ norm of the Hamiltonian $\mathcal{H}$ of the $\mu M=0.1$ BBH. We find that the use of damping does not significantly increase Hamiltonian violation.}
    \label{fig:Ham-violation}
\end{figure}

The addition of the damping term compromises diffeomorphism invariance and could, in principle, lead to constraint violations which could source significant errors to our simulations. We have checked that this is not the case in practice.  In particular, we note that the simulations with damping also exhibit a well-behaved Hamiltonian violation over time. In Fig.~\ref{fig:Ham-violation} where we show the volume-normalised Hamiltonian violation during the inspiral of the $\mu M=0.1$ BBH, we find that the scalar-damped simulation produces a Hamiltonian violation that is of the same order of magnitude as the simulation with no damping, with a slightly greater violation of approximately 10\% by $t\sim 1200M$. This suggests that no significant errors are introduced by the incorporation of the damping term. 

Despite its ability to control junk scalar radiation, the use of damping introduces two issues that require careful treatment. The first issue relates to the total energy of the spacetime. As the damping term removes energy from the system, the BH masses are observed to decrease relative to the simulations where no damping is used. This drift in the BH mass eventually settles down as the damping is gradually switched off. To account for the loss of energy due to damping and ensure a consistent comparison with the undamped simulations of GR and massless EsGB theory, we will rescale the dimensionful quantities by the measured BH masses post-damping during the inspiral\footnote{This is taken to be around $t \sim 500M$ where it is observed that the effect of damping ceases to impact the mass drift of the BHs.}. To obtain a measure of BH mass that accounts for the energy of the scalar field, we use the Wald entropy as discussed in the literature (see e.g.~\cite{Capuano:2026lhs, Julie:2019sab}), which can be expressed as
\begin{equation}
    S_W=\frac{A_H}{4}+4\pi \lambda\, \phi_H~,
\end{equation}
where $A_H$ is taken to be the measured area of the apparent horizon and $\phi_H$ the scalar field on the horizon, which we compute by averaging $\phi$ over the apparent horizon. The irreducible BH mass is related to the Wald entropy by
\begin{equation}
    M_{\rm irr}=\sqrt{\frac{S_W}{4 \pi}}\,.
\end{equation}
The total mass $M_{\rm tot}$ of the spacetime post-damping is then estimated to be the sum of the irreducible masses of the inspiralling BHs. In practice, $M_{\rm tot}$ is estimated to be $0.9860M$ and $0.9748M$ for the $\mu M=0.1, 0.2$ simulations respectively. By suitably rescaling the dimensionful quantities using $M_{\rm tot}$, we can construct dimensionless quantities, e.g. $r/M_{\rm tot}\cdot\phi$, $t/M_{\rm tot}$, that can be compared against the undamped simulations. Analogously, we perform a similar scaling for the unequal mass BBH simulation that is used later in our study, where $M_{\rm tot}$ is estimated to be $1.493M$. Note that the damping affects primarily the mass drift of the smaller BH, while the larger BH does not seem to be affected. This is to be expected, given the larger value of the scalar field at the horizon of the smaller BH. If one aims to study a system with an exact mass ratio of 1:2, one would require using ID with a slightly different mass ratio to compensate the mass loss due to damping.

The second issue is that the damping appears to introduce significant orbital eccentricity to the binary system. In order to compare the massive EsGB waveforms with those predicted by quasi-circular BBHs in massless EsGB and GR, one must reduce the eccentricity of the scalar-damped BBHs such that it approaches the quasi-circular limit. This can be achieved using standard eccentricity reduction methods of GR~\cite{Buonanno:2010yk, Habib:2024soh}, as is implemented in~\cite{Corman:2025wun} for EsGB studies.  Using the eccentricity reduced simulations, we may compare the scalar waves and gravitational waves emitted from a quasi-circular equal-mass BBH across different theories: massive EsGB, massless EsGB and GR.

The dominant channel of tensor gravitational wave emission is well-known to be the (2,2) mode (the quadrupole mode). On the other hand, the dominant radiation channel for the scalar field is usually taken to be the (1,1) mode (the dipole mode). Nevertheless, due to the special symmetry of an equal-mass BBH system, the dipolar scalar emission becomes suppressed and the dominant scalar radiation channel becomes the quadrupole mode. We therefore extract the $(2,2)$ mode radiation for both gravitational and scalar waves from our simulations. For the scalar waves, we will also extract the $(0,0)$ mode, which reflects the dynamics of the spherically symmetric part of the scalar configuration.

In Fig.~\ref{fig:BBH_SW00}, we see that there are qualitative differences in the properties of the scalar $(0,0)$ mode between a massless and massive scalar in EsGB gravity. The massive scalar (0,0) mode remains subdominant to that of the massless scalar throughout the evolution due to the suppression at large distances by the finite Compton wavelength. During the inspiral, the (0,0) mode remains approximately constant for both theories, indicating its insensitivity to the orbital dynamics. However, their behaviour deviates after the merger - while the massless scalar settles down to a different stationary value, the massive scalar is set to oscillate during the merger. It can be observed that a larger scalar mass results in higher frequency oscillations with a more suppressed amplitude. These oscillations, while appearing to be slightly damped over time, persist up to late times post-merger. Such an observation points towards the interesting possibility that having an oscillating scalar field during the inspiral might after all be permissible, if either of the progenitor BHs are formed from a previous merger (or experienced any event that triggered an oscillation in the spherically-symmetric part of the scalar field).

In the case of a massless scalar, the conventional explanation for the observed change in $\phi_{00}$ is that it reflects the change in fundamental mass scale during the evolution - as the BHs merge, the smallest curvature length scale in the spacetime increases due to the remnant BH being heavier than the individual BHs~\cite{Corman:2022xqg}. As a result, the GB invariant in the vicinity of the remnant BH becomes more suppressed after the merger, resulting in a weaker scalar configuration compared to pre-merger. The same physical mechanism should also affect the evolution of the massive scalar. However, this does not seem to be the dominant factor in determining the late-time behaviour of $\phi_{00}$, which is instead determined by the oscillations induced during the merger. This highlights the scalar mass's role in trapping long-lived modes in the scalar field, and its potential in contributing to new late-time dynamics that are absent in the massless case.

The scalar $(2,2)$ mode, reflecting the orbital dynamics' influence on the scalar field, is shown in Fig.~\ref{fig:BBH_SW22}. Contrary to the massless case where the $(2,2)$ mode smoothly increases in frequency and amplitude during the inspiral towards the merger, the massive scalar waves of this mode are relatively suppressed during the inspiral phase. In fact, this can be seen as another manifestation of the confining nature of the scalar mass. If one extracts the $(2,2)$ mode radiation at a smaller extraction radius, one finds that the radiation is in fact non-zero\footnote{While a formal extraction of scalar/gravitational radiation should be carried out at infinity, this comparison nevertheless illustrates that non-trivial dynamics can arise locally in a massive scalar theory.} (see Fig.~\ref{fig:ext_rad_comparison}). However, the frequency of the inspiral is likely too low to generate quadrupolar scalar radiation that can escape the massive potential to infinity. As a result, the quadrupolar scalar emission is suppressed by the mass during the inspiral. As the system approaches merger, the frequency of the quadrupolar scalar waves becomes high enough for the radiation to escape the massive barrier. This explains our observation that at large extraction radius, one only sees the burst of radiation near merger. We note that similar observations have been speculated on in studies of massive scalar-tensor gravity~\cite{PhysRevD.108.064057}, and analogous conclusions can be found in perturbative studies of such theories~\cite{Alsing:2011er, Diedrichs:2023foj}, where the scalar mass sets a minimum inspiral frequency for scalar waves to be emitted.  After the merger, the scalar $(2,2)$ mode appears to undergo a ringdown phase, followed by oscillations at the scalar mass frequency before eventually decaying to noise level. We will analyse the post-merger signal in more detail later in this section. 

Another feature we can notice in Fig.~\ref{fig:ext_rad_comparison} is the apparent extraction radius-dependent morphology of the merger waveform. This is perhaps unsurprising given the dispersive nature of massive scalar waves, where different frequency components of the wave propagate at different speeds. Similar discussions can be found in studies of massive scalar-tensor theories~\cite{OlveraM:2026mah}. This highlights the fact that analysing massive scalar waves extracted at a finite extraction radius can lead to extraction radius-dependent conclusions. 

Finally, the gravitational waves (2,2) mode is represented in Fig.~\ref{fig:BBH_GW22}, comparing the massless and massive EsGB waveforms with those of GR. We note the difficulty in comparing the dephasing of the massive EsGB waveforms relative to the other two waveforms during the inspiral, as is commonly done in beyond-GR studies~\cite{Julie:2024fwy,Corman:2025wun,Capuano:2026lhs,Lara:2026xcs}. This is due to the fact that the remaining eccentricity in our massive EsGB BBHs effectively changes the system and makes it difficult to perform a consistent alignment in the early inspiral against the undamped simulations. While this can in principle be improved with further eccentricity reduction iterations, such a precise treatment of eccentricity requires heavy computational costs and lies outside of the scope of our current work \footnote{We note that effort towards understanding dephasing in massive EsGB is currently underway \cite{Huetal}.}. Nevertheless, if one aligns the waveforms at the merger as shown in Fig.~\ref{fig:BBH_GW22}, we find the expected behaviour that they exhibit only minor deviations from one another.

\subsection{Ringdown}\label{sec:ringdown}

We now turn our attention to the ringdown part of our waveforms and attempt to extract some physical predictions. Recent work has shown that new fields that couple non-minimally to gravity, as is the case in the scenario we are simulating here, modify the GW frequency spectrum not only by shifting the GR-predicted quasi-normal mode (QNM) frequencies, but also ``contaminate'' the signal with additional frequencies that correspond to the QNM spectrum of the new field~\cite{Lestingi:2025jyb,Asuncion:2026efk}. This had previously been observed in specific theories~\cite{Molina:2010fb,Blazquez-Salcedo:2016enn,Witek:2018dmd} and considered in the context of modelling ringdowns~\cite{Crescimbeni:2024sam}. These results motivate us to study the frequency content of the post-merger gravitational and scalar modes.

We will focus primarily on frequency contamination effects. These effects will arise as independent peaks in the frequency spectrum, thereby serving as a smoking gun beyond-GR signature. Identifying frequency shifts relative to GR is far more challenging, as they are very small for the values of the coupling constant we are considering (which is well-motivated by EFT considerations), while the accuracy of our simulations and frequency extraction methods is limited. To understand the physical origin of the extracted frequencies, we will compare the extracted values to the readily computed QNMs of a Kerr BH in GR (see e.g.~\cite{Chung:2024vaf, Blazquez-Salcedo:2024oek, Khoo:2024agm, Blazquez-Salcedo:2017txk, Pierini:2022eim, Lui:2026osx} for recent developments in 
computing QNMs in EsGB gravity). Note that one can also in principle study the excitation amplitudes and phases of QNMs~\cite{Hu:2025bkg}. However, due to its stronger dependence on the progenitor system where eccentricity can play an important role, we will postpone such an analysis to a future work. Our analysis of the ringdown mode content is partially motivated by and complementary to the study of ringdown contamination for massive fields pursued in Ref.~\cite{Daddarioetal} in the context of black hole perturbation theory.

\begin{figure}
    \centering
    \includegraphics[width=1\linewidth]{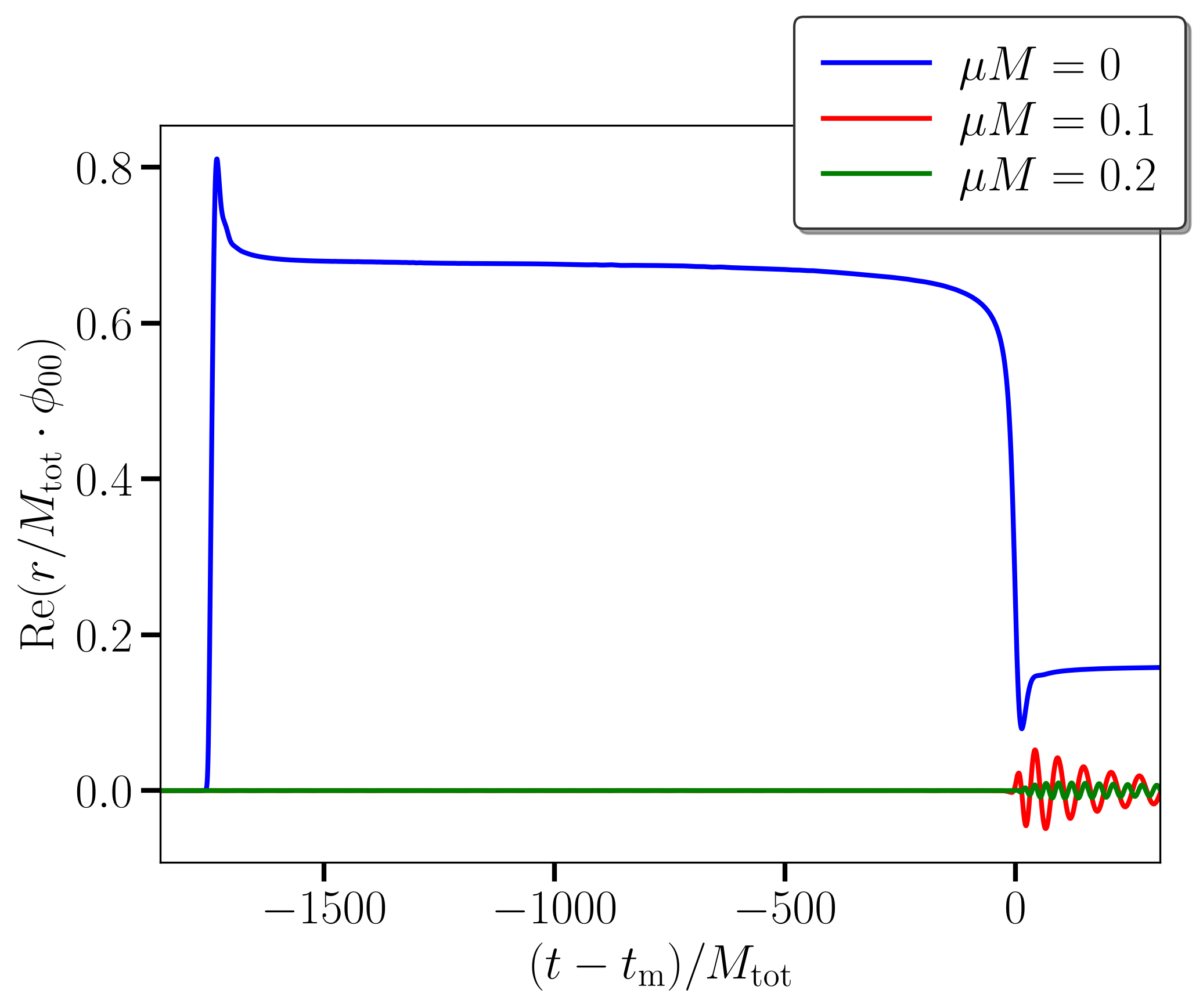}
    \caption{$\phi_{00}$ mode extracted at $r=100M$ plotted as a function of time, where $t_{\rm m}$ is the merger time. While the massless scalar settles from one stationary value to another after merger, the massive scalars exhibit an oscillatory behaviour post-merger.}
    \label{fig:BBH_SW00}
\end{figure}

\begin{figure}
    \centering
    \includegraphics[width=1\linewidth]{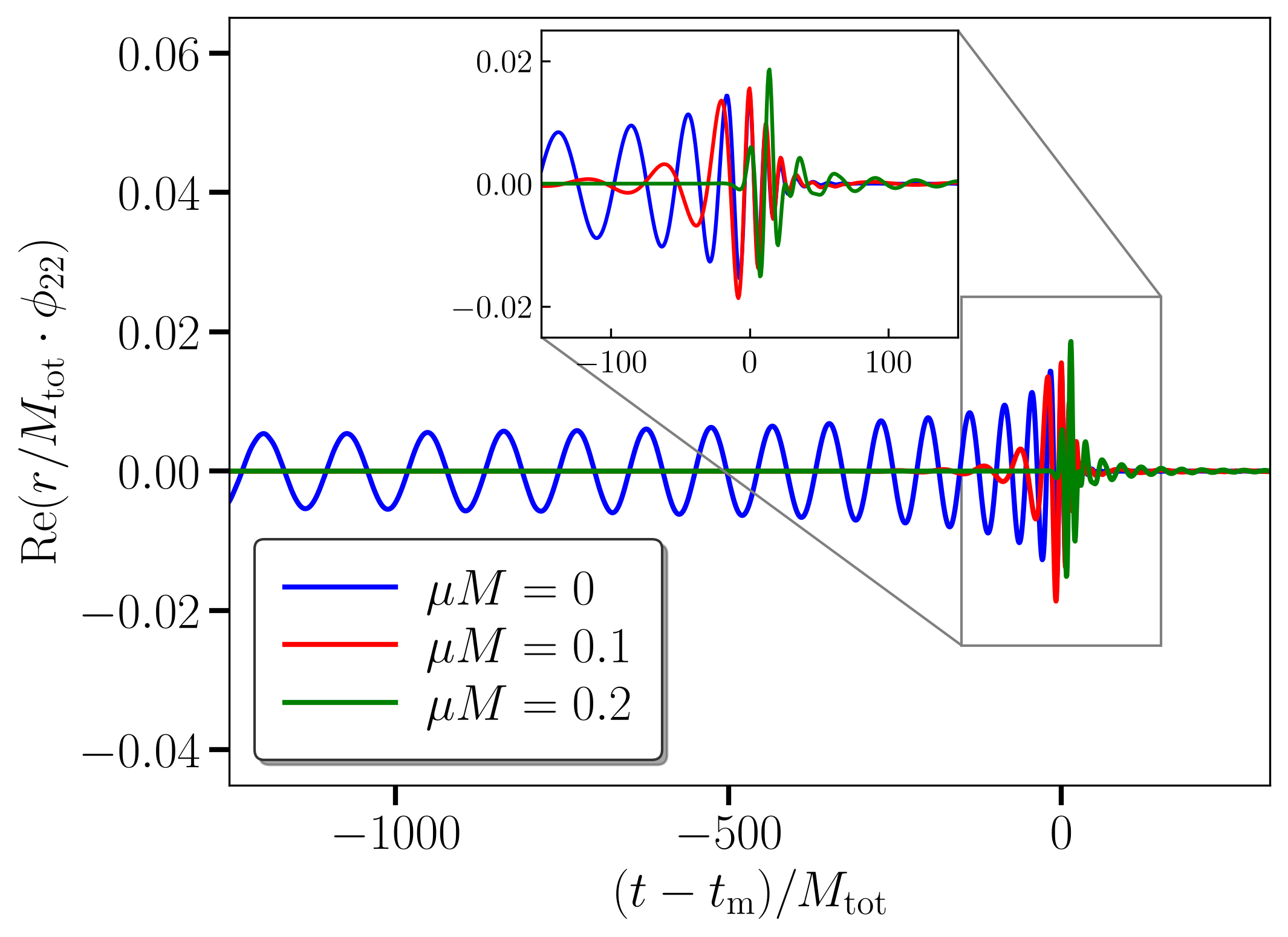}
    \caption{$\phi_{22}$ mode extracted at $r=100M$ plotted as a function of time. The massless and massive scalars exhibit a different phenomenology: the former continuously emits scalar radiation during the inspiral, whereas the same emission from massive scalars are quenched until near merger.}
    \label{fig:BBH_SW22}
\end{figure}

\begin{figure}
    \centering
    \includegraphics[width=1\linewidth]{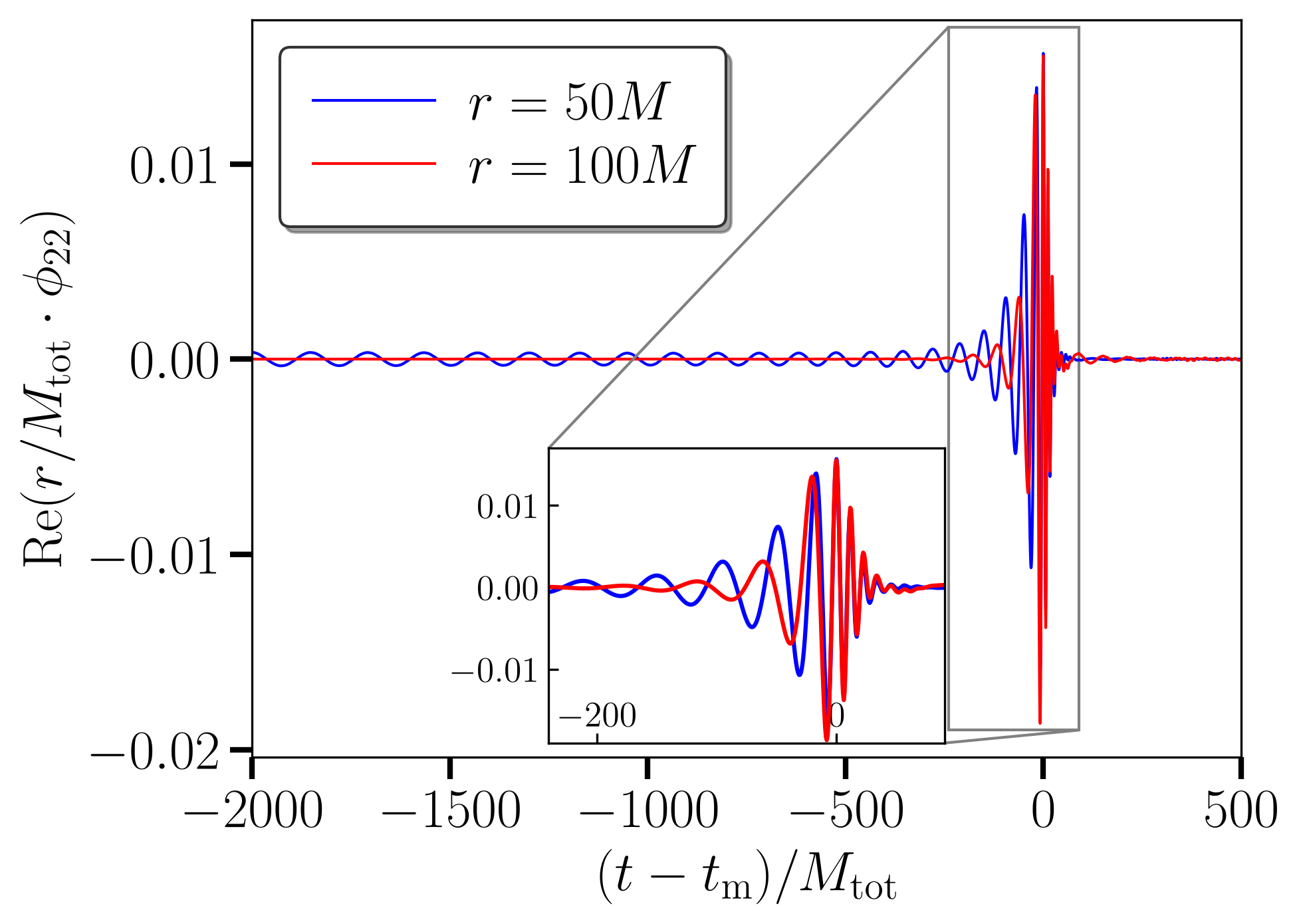}
    \caption{Comparing $\phi_{22}$ for a $\mu M=0.1$ scalar field as extracted at $r=50M$ and $r=100M$ respectively. While the scalar field does not exhibit interesting dynamics in the far-field region during most of the inspiral phase (as reflected by the red curve), it shows non-trivial dynamics when extracted at a closer radius (e.g. $r=50M$) to the BBH. This demonstrates that massive scalar dynamics in the strong field do not necessarily result in scalar radiation to infinity.}
    \label{fig:ext_rad_comparison}
\end{figure}

\begin{figure}
    \centering
    \includegraphics[width=1\linewidth]{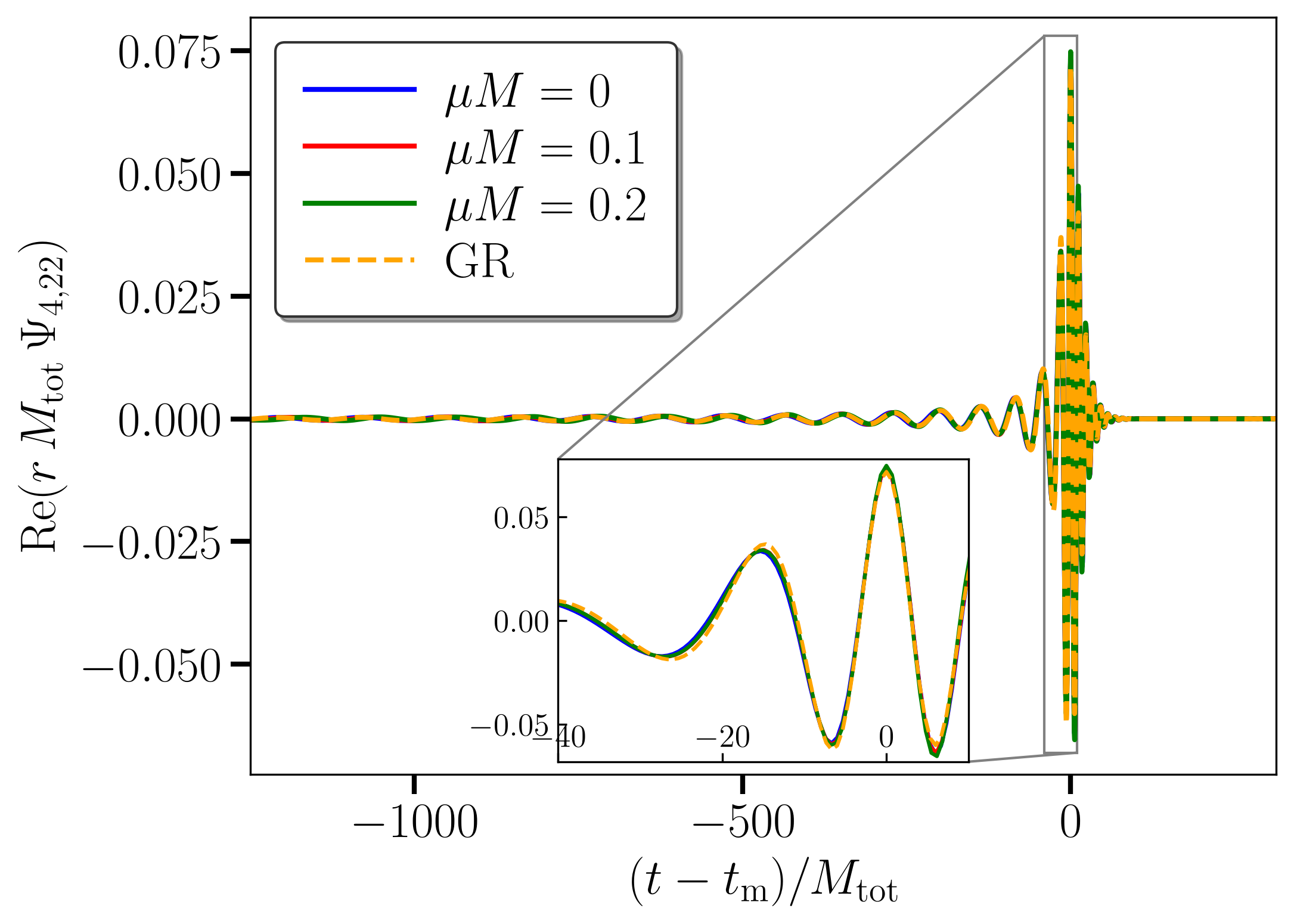}
    \caption{$\Psi_{4,22}$ mode extracted at $r=100M$ plotted as a function of time. While a frequency-alignment of the waveforms during early inspiral is not possible due to the residual eccentricity in the damped simulations, we nevertheless show that all waveforms exhibit a similar structure near the merger, as demonstrated in the inset.}
    \label{fig:BBH_GW22}
\end{figure}

Before presenting our results, we first outline our strategy.  For the ringdown of the gravitational waves $(2,2)$ mode, it is well-known that the signal behaves as an exponentially decaying sinusoid. As such, one may perform a time domain fitting to extract the frequencies of the ringdown modes. In particular, we use the non-linear model
\begin{equation}
    y=\sum_{i=1}^{n}e^{-\gamma_it}(A_i\cos{\omega_it}+B_i\sin{\omega_it})+C\,,
\end{equation}
where $n$ is taken to be $1$ for a one-mode fit and $2$ for a two-mode fit (to check for the possibility that the ringdown contains a scalar frequency contamination). Whether a secondary frequency exists can be deduced by checking the stability of the fitted parameters as we vary the fitting window.

\begin{figure}
    \centering
    \includegraphics[width=1\linewidth]{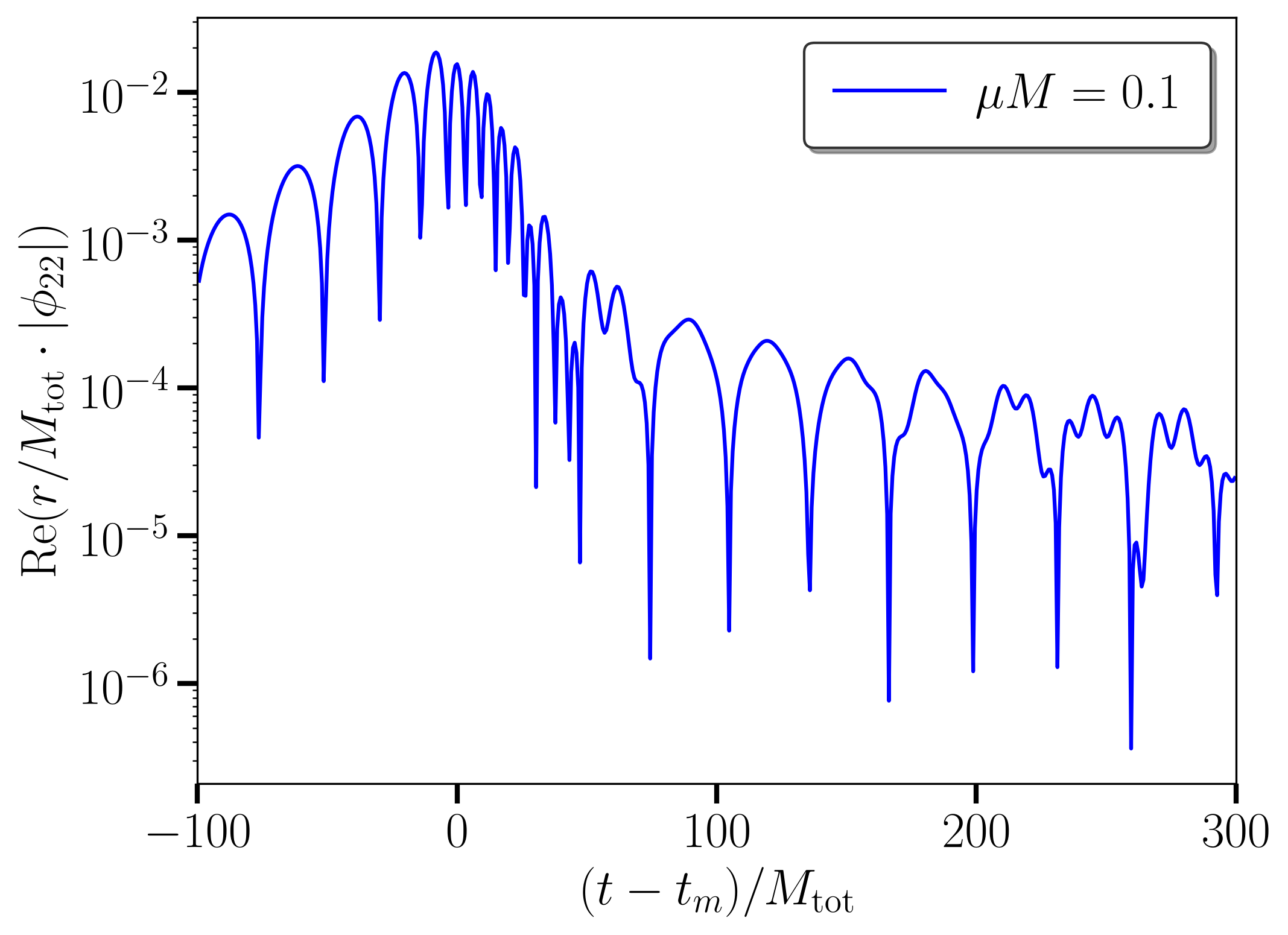}
    \caption{The near merger behaviour of $\phi_{22}$ for the $\mu M=0.1$ BBH, plotted in log-scale. We note that the ringdown does not appear to be sufficiently described by a sum of exponentially damped sinusoids, which motivates the use of frequency domain analysis.}
    \label{fig:log_sw22}
\end{figure}

Although this method works well for gravitational waves, it is not immediately obvious whether it can be applied to scalar waves. Indeed, the post-merger time domain scalar waveforms do not appear to be adequately described by a collection of exponentially decaying sinusoids (see Fig.~\ref{fig:log_sw22} for an example). This motivates us to consider a frequency domain analysis of scalar waves using Fast Fourier Transforms (FFT), which produces a Fourier spectrum that highlights the dominant frequency contributions in the ringdown. To mitigate the effects of spectral leakage, we supplement our FFT analysis with a separate analysis using the Welch's method.\footnote{The Welch's method~\cite{welch1967use} estimates the power spectral density (PSD) by dividing the data into shorter segments and computing the periodogram for each segment using a window function, before averaging these periodograms to produce the full PSD.}  The Welch PSD then provides an alternative tool for estimating frequency peaks. To obtain a more accurate estimate of the dominant frequencies from the peaks of our discrete spectrum/PSD, we implement a $3-$point Gaussian interpolation around the frequency bins that locally maximises the amplitude~\cite{Gasior:2004md}. The final issue we have to address is the choice of fitting windows, which is made delicate due to the short duration of the ringdown. We will discuss our strategy for choosing fitting windows in Appendix \ref{sec:fittingwindows}, where we also evaluate the reliability of Fourier-based methods.

We first analyse the gravitational wave $(2,2)$ mode ringdown data of GR and massless/massive EsGB simulations, using the time-domain fitting approach outlined above. The results are presented in Table \ref{tab:GW22}. Across all the cases considered, we find that the two-mode fitting of the time-domain waveform does not yield a stable secondary frequency upon changing fitting windows, indicating the presence of only one dominant frequency mode in the data. On the other hand, the frequency extracted from the one-mode analytical fit agrees well with the expected QNM value of $M_{\rm f}\cdot\omega=0.524$, where $M_{\rm f}$ is the mass of the remnant BH, for a spin-2 perturbation of a Kerr BH in GR with dimensionless spin of 0.68~\cite{Berti:2005ys, Berti:2009kk}. The values of $M_{\rm f}$ and dimensionless spin are measured at the apparent horizon of the remnant BH. Note that there is an approximate 2\% difference between the extracted QNM frequencies of GR and EsGB with $\mu M=0.2$. As discussed above, we cannot conclude with strong confidence within the accuracy of our methods whether the observed frequency shift is a genuine physical effect or an artefact of our fitting methods. At the same time, we see no evidence of scalar contamination in the form of an extra frequency component in the gravitational wave $(2,2)$ mode ringdown.

\begin{table}[ht]
\centering
\begin{tabular}{||c|c||}
\hline
 & $M_{\rm f}\cdot\omega_{1 \rm mode}$ \\
\hline
GR   & 0.525 \\
$~\mu M=0~~$   & 0.526 \\
$~\mu M=0.1~~$ & 0.530 \\
$~\mu M=0.2~~$ & 0.536 \\
\hline
Expected value (GR) & 0.524\\
\hline
\end{tabular}
\caption{Extracted frequencies for the ringdown of $(2,2)$ mode gravitational waves in GR, massless EsGB and massive EsGB respectively. The frequencies are extracted from a one-mode fitting of the time-domain waveform ($M_{\rm f}\cdot\omega_{1 \rm mode}$), where a strong agreement with the expected value of QNM is found.}
\label{tab:GW22}
\end{table}

We now turn to the analysis of the scalar ringdown using Fourier-based methods. For the scalar waves $(0,0)$ mode, no clear ringdown in the traditional sense (i.e. an exponentially decaying interval) can be found in the waveform. This is perhaps unsurprising, given that the $(0,0)$ mode of the massless scalar waves in linear EsGB theory also does not exhibit a clear ringdown (as is evident from Fig.~\ref{fig:BBH_SW00}). However, unlike the massless case, the massive waves do not settle to a stationary value. Instead, the oscillations triggered at merger persist till late times. We perform a spectral analysis on the post-merger data to understand the frequency content of these oscillations. The frequency spectrum/PSD reveals that there is only one dominant frequency corresponding to the scalar mass (with a measured frequency of $M\cdot\omega \approx0.101$ for $\mu M=0.1$ and $M\cdot\omega \approx0.203$ for $\mu M=0.2$), as is evident from the time domain representation of the waveform. The lack of other frequency peaks corresponding to the gravitational sector indicates that the scalar (0,0) mode does not couple strongly with gravitational radiation.

\begin{figure}
    \centering
    \includegraphics[width=1\linewidth]{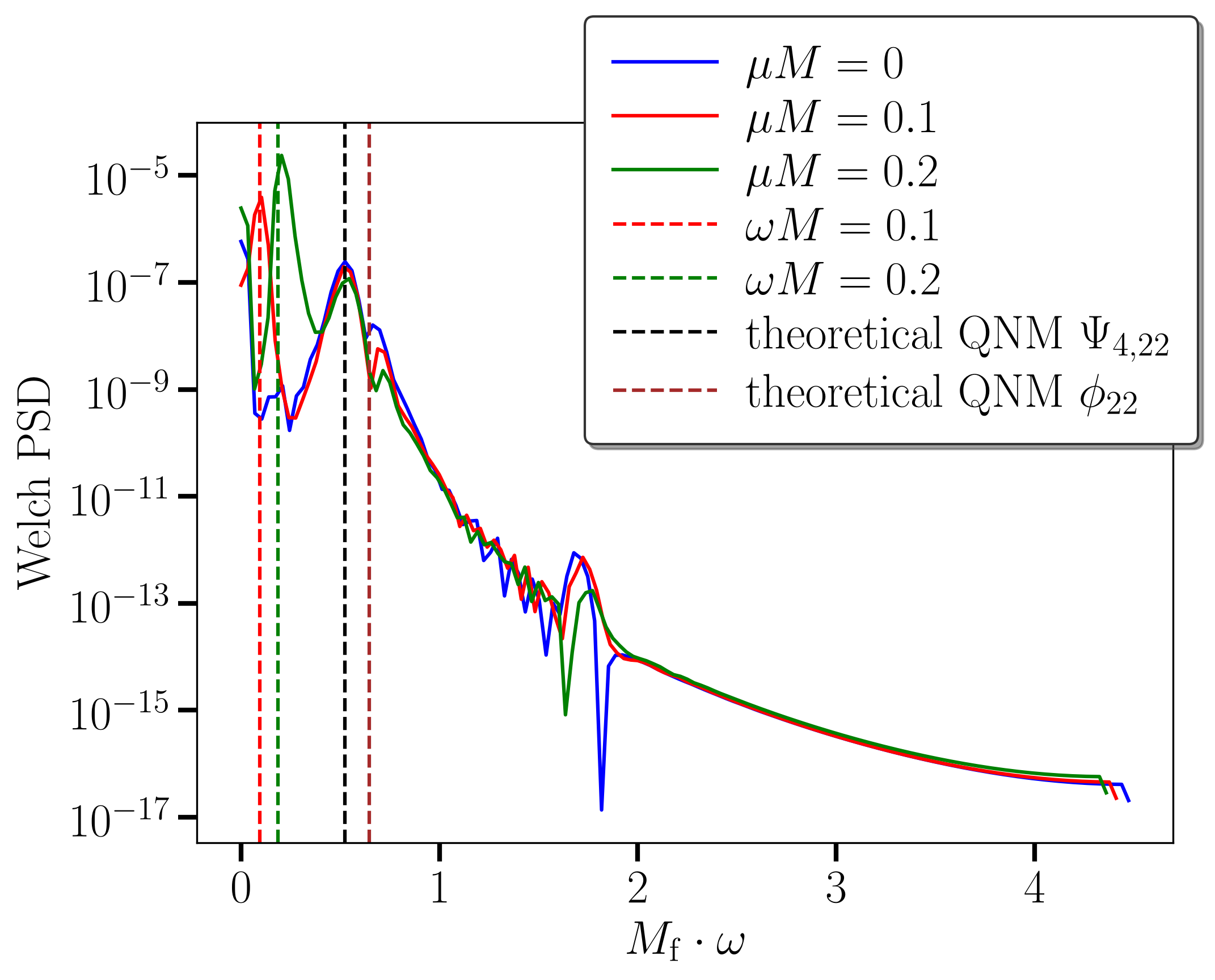}
    \caption{The Welch PSD of the $\phi_{22}$ ringdown as a function of dimensionless frequency $M_{\rm f}\cdot\omega$ from the equal-mass BBH mergers. The three PSD's correspond to the three scalar masses: $\mu M=0$ (blue), $\mu M=0.1$ (red), and $\mu M=0.2$ (green). The vertical lines are the expected frequencies from different physical origins: the red and green dotted lines correspond to the scalar mass frequencies of $\omega M=0.1, 0.2$ respectively, the black dotted line corresponds to the expected QNM frequency for a spin $s=-2$ perturbation of a Kerr BH, and the brown dotted line corresponds to the expected QNM frequency for a spin $s=0$ perturbation of a Kerr BH.}
    \label{fig:psd_SW22}
\end{figure}
For the scalar waves (2,2) mode, the post-merger signal in the time domain differs qualitatively between the massless and massive cases. While the massless waveform exhibits a clean ringdown until noise dominates the signal, the massive ringdown appears to initially consist of a mixture of frequencies, followed by a stage where the signal is dominated by a single frequency, before eventually reaching the noise level. To understand the frequency contributions to the ringdown of the scalar (2,2) mode, we analyse its PSD as shown in Fig.~\ref{fig:psd_SW22}, which compares the three scalar masses of $\mu M=0,0.1$ and 0.2 (see Table \ref{tab:SW22} for the extracted frequencies from the spectrum/PSD). For massive scalar waves, we can immediately identify a frequency component that is associated with the scalar mass (labelled by the red and green dotted lines for $\mu M=0.1, 0.2$ respectively), indicating that the scalar mass can also leave an imprint on the asymmetric scalar radiation in addition to the spherically symmetric (0,0) mode. Furthermore, we found a common frequency peak for all three cases around $M_{\rm f}\cdot\omega\sim0.53$. This frequency agrees with the expected QNM frequency of $M_{\rm f}\cdot\omega=0.524$ for a (2,2) mode spin-2 perturbation (labelled as the black dotted line), which suggests that it is the contamination of scalar modes by spin-2 frequencies originating from the scalar's coupling to the tensorial gravitational waves. Given that the Gauss-Bonnet invariant serves as a source in the scalar field's equations of motion, it is expected that the geometric degrees of freedom can leave an impact in the scalar's dynamics, which in this case manifests as an extra frequency component in the ringdown of the scalar (2,2) mode.

It is interesting that we observe contamination in the scalar $(2,2)$ mode while we do not see any evidence of it in the gravitational (2,2) mode. While the effect is a result of coupling between the modes, which is in turn an outcome of the non-minimal coupling between the scalar and the metric, the amount of contamination in each direction generally depends on the relative amplitude between metric and scalar excitations~\cite{Lestingi:2025jyb}. As is evident from Figs.~\ref{fig:BBH_SW22} and \ref{fig:BBH_GW22}, the maximum dimensionless amplitude of the scalar waves at the merger is approximately a factor of 3 or 4 weaker than that of the gravitational waves in our simulations. This strongly suggests that the particular system we are studying - an equal-mass, low-eccentricity binary of non-spinning BHs - in this particular theory does not excite the scalar field sufficiently for it to leave an observable contamination imprint on the gravitational ringdown waveform. Whether a more general system (e.g. large mass asymmetry, eccentricities, spins or precession, etc.) can do better in this regard in this or another theory will be revisited in a future work.

Finally, we observe one subdominant peak that neighbours the spin-2 frequency peak across all three scalar masses considered. We note that this frequency can only be found in the Welch PSD if we use a fitting window that starts at a sufficiently early time ($t\sim10M$). Even then, this peak is never clearly visible from the FFT spectrum, as it is likely buried within the wider peak associated with the spin-2 mode. Given its relatively weak amplitude and high sensitivity to our spectral analysis methods, its interpretation becomes more subtle compared to the previous peaks we extracted. Nevertheless, due to its close resemblance to the expected QNM frequency of $M_{\rm f}\cdot\omega=0.648$ associated with a spin-0 perturbation of a Kerr BH, we postulate that it can be physically interpreted as the frequency associated with the free decay of the scalar field on a spinning BH background~\cite{Witek:2018dmd}. In this case, we would not expect a close agreement between the massive scalar frequencies to the expected value for massless scalar perturbations (see e.g.~\cite{Alves:2025erh, Lui:2026osx}), but the latter should better coincide with the extracted massless scalar frequency. This is indeed supported by the observations from the Welch PSD presented.

\begin{figure}
    \centering
    \includegraphics[width=1\linewidth]{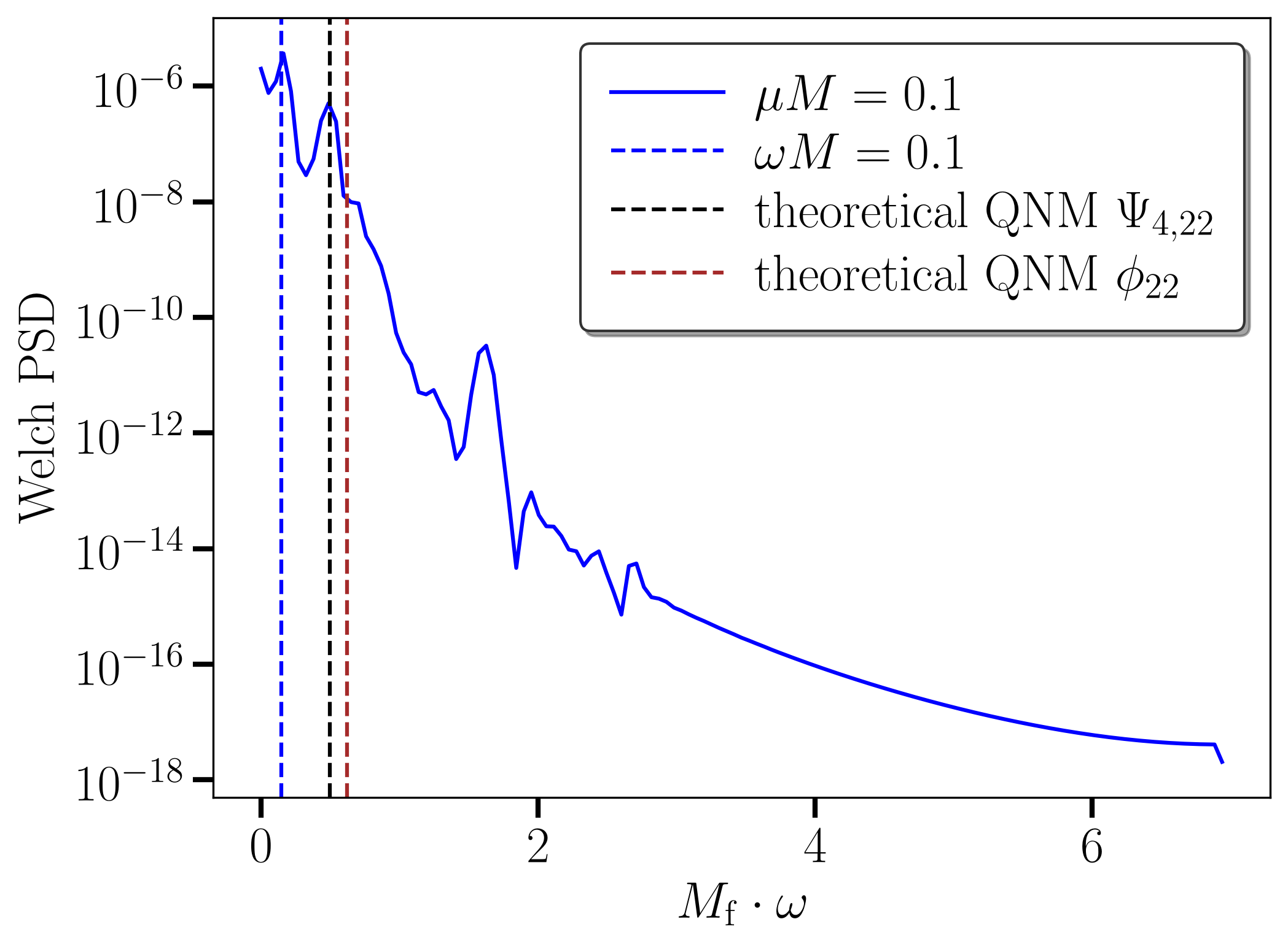}
    \caption{The Welch PSD of the $\phi_{22}$ ringdown as a function of dimensionless frequency $M_{\rm f}\cdot\omega$ from the unequal mass BBH merger, showing qualitative agreement with the results from equal-mass BBH mergers that the waveform receives contributions from the (i) scalar mass frequency, the (ii) scalar-tensor coupling, and the plausible (iii) free decay frequency of the scalar (2,2) mode.}
    \label{fig:psd_unequal_sw22}
\end{figure}

As an independent verification of the existence of the third peak, we performed one simulation of an unequal mass BBH with mass ratio 1:2 using a scalar mass of $\mu M=0.1$. The remnant BH has a different dimensionless spin ($\sim 0.61$) from our equal-mass BBHs, providing an alternative setting to verify the physical origin of the extracted frequency. The post-merger PSD of the scalar $(2,2)$ mode once again produced two dominant frequency peaks at $M \cdot\omega \approx0.110$ (the scalar mass frequency) and $M_{\rm f} \cdot\omega \approx 0.483$ respectively, along with a third plausible peak at $M_{\rm f} \cdot\omega \approx0.651$ (see Fig.~\ref{fig:psd_unequal_sw22}). The second peak is in strong agreement with the expected QNM of $M_{\rm f}\cdot\omega=0.497$ for a spin-2 perturbation of a Kerr BH with our measured spin, while the third peak is in good agreement with the expected QNM frequency of $M \cdot\omega = 0.621$ for a spin-0 perturbation. This supports our conclusions from the equal-mass BBH simulations that the scalar (2,2) mode consists of frequencies arising from (i) the scalar mass, (ii) the scalar-tensor coupling, and a weaker contribution from (iii) the free decay frequency of the scalar (2,2) mode.  Using the unequal mass BBH simulations, we have also re-attempted our search for scalar contamination in the gravitational (2,2) mode radiation.  Our conclusions remain in agreement with the equal-mass case, in which only one frequency mode is present and no frequency contamination is identified.

\begin{figure}
    \centering
    \includegraphics[width=1\linewidth]{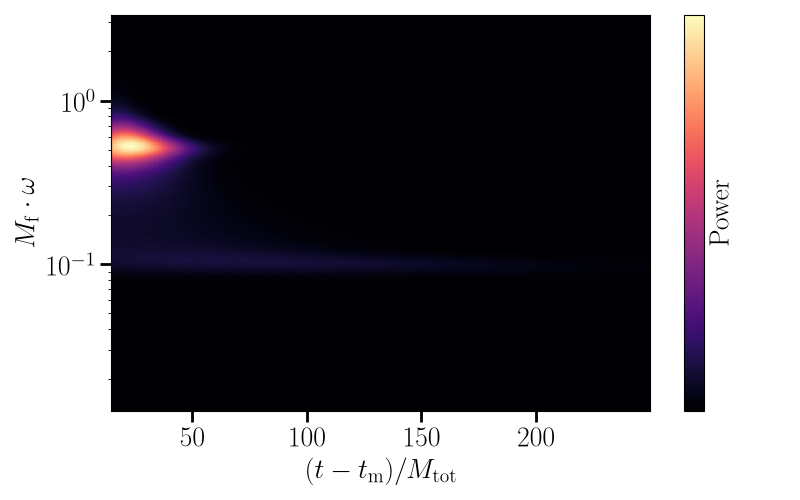}
    \caption{Scalogram produced by the Continuous Wavelet Transform of the scalar (2,2) mode ringdown of a massive scalar with $\mu M =0.1$.
    }
    \label{fig:scalogram}
\end{figure}

Although the PSD provides information on the frequency content of the ringdown signal, it does not reveal when a particular frequency component arises within the signal. To recover such temporal information, we can analyse the scalogram produced by performing a Continuous Wavelet Transform\footnote{The Continuous Wavelet Transform estimates the frequency content of a signal by fitting it against a function (the ``wavelet'') that can be suitably scaled to detect a range of frequencies. This produces a scalogram, which shows the frequency content of the signal as a function of time.}~\cite{10.1007/978-3-642-75988-8_1}, which we illustrate in Fig.~\ref{fig:scalogram} for the $\mu M=0.1$ scalar (2,2) mode ringdown of the equal-mass BBH remnant. Early in the ringdown, the scalar (2,2) mode is dominated by frequencies near $M_{\rm f}\cdot\omega \sim 0.5$, which is approximately the range probed by the spin-2 and free scalar QNM frequencies. Note that the two modes cannot be distinguished from the scalogram due to the wide frequency spread. While the scalar mass frequency ($\mu M =0.1$ for the results presented here) is excited early during the ringdown, it does not dominate the signal until the $M_{\rm f}\cdot\omega \sim 0.5$ frequencies have sufficiently decayed. This observation aligns with our interpretation from the time domain waveform that the early ringdown is composed of a mixture of frequencies, followed by a period where the signal is dominated by one frequency, before eventually decaying to noise.

\renewcommand{\thempfootnote}{\arabic{mpfootnote}}
\begin{table}[ht]
\addtocounter{footnote}{1}
\xdef\mytablefn{\arabic{footnote}}
\centering
\begin{tabular}{c c c c}
\hline
 & $M\cdot\omega_1$ & $M_{\rm f}\cdot\omega_2$ & $M_{\rm f}\cdot\omega_3$\\
\hline
$\mu M=0$ & / & 0.526 & 0.674 \\
$\mu M=0.1$ & 0.108 & 0.522 & 0.704 \\
$\mu M=0.2$ & 0.224 & 0.533 & 0.722 \\
\hline
Expected values (GR) & $\mu M$ & 0.524 & 0.648 \\
\hline
\end{tabular}
\caption{Extracted frequency peaks for the (2,2) mode scalar waves in the ringdown of an equal-mass BBH merger. For the massive scalar waves, we find frequency contributions arising from the i) scalar mass, ii) spin-2 contamination, and iii) the free decay of the scalar field; while the massless scalar only contains contributions from the latter two\protect\hyperlink{tab:SW22fn}{\textsuperscript{\mytablefn}}. }
\label{tab:SW22}
\end{table}
\footnotetext[\mytablefn]{\hypertarget{tab:SW22fn}{}Note that $M_{\rm f}\omega_3$ can only be identified using the Welch PSD with a fitting window that includes sufficiently early ringdown data. Only one fitting window we used matches this criterion, implying that unlike the other values presented, its measured value is not averaged over different fitting windows and spectral analysis methods we used.}

To summarise, while the scalar waves ringdown (in particular the (2,2) mode) exhibits rich features including free scalar dynamics and signatures of scalar-geometry coupling, the gravitational (2,2) mode ringdown appears less sensitive to the scalar dynamics. Nevertheless, there remains one possibility where we can observe scalar-geometry coupling in the gravitational waves. From the massive EsGB simulations, we find that the scalar field dynamics persists up to late times instead of settling down to a stationary state. One is then led to question whether such remnant dynamics in the scalar field is capable of sourcing gravitational waves at very late times by leaking energy to the quadrupolar radiation channel via non-linearities in the equations, resulting in analogous observations in massive Chern-Simons gravity where quasi-bound state frequencies of the scalar contaminate the gravitational perturbations~\cite{Macedo:2018txb}. To answer this question, we extended our simulations to late times after merger ($t-t_m \sim 1000M$) and searched for a continuous frequency component in the scalogram of the gravitational waves, with the expectation that, if such a coupling does exist, it would manifest itself as a continuous emission of gravitational waves at a fixed frequency. However, it appears that the late time data is dominated by noise, which limits our ability to confidently identify the presence of scalar-induced gravitational waves. We therefore postpone a discussion of the potential late time scalar-induced gravitational waves to a future work.

We conclude this section by discussing some limitations of our ringdown analysis. The short duration of the ringdown (relative to the period of QNM oscillations) fundamentally limits the accuracy of Fourier-based methods in frequency extraction, due to the limited cycles of oscillations one can include in a given fitting window. This also leaves fewer room for changing the fitting windows we can use in our analysis (especially with our strategy in choosing fitting windows as outlined in Appendix \ref{sec:fittingwindows}), since using a window that is too short will result in poor frequency resolution. Moreover, the broad and occasionally distorted peaks observed in the Fourier spectrum (likely the combined result of having a finite QNM decay time and spectral leakage) also make peak frequency extraction using Gaussian interpolation less reliable. Nevertheless, the preliminary analysis above still provides a qualitative guide into the frequency content of the ringdown of the different waveforms considered and serves as a proof of principle that physical modes of various origins can be identified in the ringdown by analysing its frequency components.

\section{Discussion}
We have studied BBH mergers in massive EsGB gravity, building on existing literature where a massless scalar is usually assumed. To determine the dynamics of a BBH merger, one must provide suitable constraint-satisfying initial data for all relevant dynamical fields. However, an initial data solver for EsGB gravity is not yet available and the standard approach when considering a massless scalar field has been to take the field and its conjugate momentum to vanish initially. The EsGB constraints then reduce to their GR counterparts, allowing one to make use of GR initial data for the metric to initialise the simulations, at the expense of generating junk scalar radiation at the start of the simulation. This junk radiation leaves the system sufficiently fast in studies involving a massless scalar, so as to not become a concern.

Our simulations demonstrate clearly that this is not the case for a massive scalar. Instead, the junk radiation lingers in the numerical domain until late times, thereby making the simulation sensitive to initial data artefacts and unusable from a physics pespective. We have introduced a new approach that can overcome this problem, which involves modifying the EoM of the scalar field by adding a damping term and gradually switching off the damping strength. We have shown that the junk dynamics in the scalar field can be efficiently tamed in this way. In the case of an isolated BH, the post-damping asymptotic behaviour of the scalar field agrees with the analytical approximation. As for a BBH merger, we found no significant new junk scalar waves being triggered when switching off the damping term after a sufficiently long relaxation time. These results strongly indicate that the post-damping scalar configuration is sufficiently close to the quasi-stationary one admitted by the theory.

A side-effect of our method of taming junk radiation when applied to BBHs is that the binary acquires significant eccentricity during the relaxation period. We have regulated that by implementing eccentricity reduction schemes, which allowed us to bring the BBHs closer to a quasi-circular orbit. Using these simulations, we have then performed a qualitative study of waveforms of each scalar mass. Firstly, the behaviour of the scalar (0,0) mode radiation exhibits significant differences when comparing the massless and massive scalar case. Although the (0,0) mode appears to be relatively constant during the inspiral phase in both cases, the post-merger waveform shows an oscillatory profile for a massive scalar, while the massless scalar simply settles down to a different stationary value. On the other hand, from the scalar (2,2) mode waveform, we see that the scalar waves with frequency below the scalar mass threshold do not propagate to large distances. Once the inspiral frequency exceeds the scalar mass threshold, the scalar radiation channel becomes more effective in radiating energy away from the binary. Despite these scalar mass-dependent differences in the scalar waveforms, we find no significant deviations in the gravitational $\Psi_{4,22}$ waveform between all theories considered. All of the above are consistent with theoretical expectations for an equal-mass binary, which is the case we consider, and hence demonstrate that our junk radiation taming method is effective.

Finally, we have used
 the ringdown portion of the scalar and gravitational waves to perform a preliminary analysis of the frequency content in the different waveform modes. The frequency associated with the mass of the scalar features clearly in the scalar ringdown and affects the waveform morphology. Our ringdown frequency analysis is partially motivated by recent work showing that in theories such as the one we consider the ringdown signal should exhibit frequency contamination \cite{Lestingi:2025jyb}. It is also motivated by the expectation that the coupling between scalar and metric perturbation that leads to contamination will affect the ringdown more strongly in the case of massive fields \cite{Daddarioetal}. We do not find evidence of scalar contamination in gravitational wave signal but we do identify gravitational contamination in the scalar waves. This asymmetry in the effect appears to be due to the low relative amplitude between scalar and metric perturbations, which is known to control the size of contamination \cite{Lestingi:2025jyb}. This amplitude ratio is in turn related to the morphology of the binary, and the amount of non-linearity in the scalar sector of the theory. It would hence be particularly interesting to explore different theories and binary configurations in the future.

Our results demonstrate that the scalar field remains dynamical after the merger instead of settling down to a stationary state, as it keeps oscillating at the scalar mass frequency (most notably in the (0,0) mode). At the same time we also see clear evidence of the coupling between the scalar and tensor modes. An intriguing future direction is to explore whether the combination of these two effects can lead to an observable signature in the late ringdown, when the BHs would have already settled to its stationary state in GR. This is being explored using BH perturbation theory in~\cite{Daddarioetal} and it would also be interesting to explore numerically. This would require improvements to our code that would allow it to reach the required level of numerical accuracy for an extended period of time after the merger.

To conclude, it is worth emphasising that the damping method we propose in this work should be viewed as a workaround, rather than a final solution, to the junk radiation problem for massive fields. It can allow one to draw some physical insights in the absence of an initial data solver, but our results also emphasise the importance of developing such a solver for modified gravity. Indeed, it is unclear and worth exploring in the future how well the damping method will work for massive EsGB in other astrophysically relevant scenarios such as a binary neutron star mergers, or in theories where higher order self-interactions of the scalar are considered. At the same time, it is also worth noting that some junk radiation will inevitably be present even when using constraint-satisfying initial data and it will linger on when massive fields are present. Therefore the damping method we developed could prove useful alongside a well-developed initial data solver.

\section{Acknowledgements}
We acknowledge helpful conversations with Katy Clough, Giovanni D'Addario, Nicola Franchini, Herry F. Lalus, Jacopo Lestingi and Sebastian H. V\"olkel. 
We thank Maxence Corman for sharing the scripts to perform eccentricity reduction, and Giovanni D'Addario and Daniela Doneva for providing useful comments on the manuscript. 
HLHS is supported by the Croucher Scholarship and the School of Mathematical Sciences, University of Nottingham Postgraduate Studentship.
In the early stages of this work, LAS was supported by a London Mathematical Society (LMS) Early Career Fellowship. LAS has also been partly funded by Interuniversitaire Bijzonder Onderzoeksfonds (IBOF)/21/084.
MB acknowledges partial support from the STFC Consolidated Grant nos. ST/Z000424/1, UKRI2492 and ST/C004235/1.
TPS acknowledges partial support from the STFC Consolidated Grant nos. ST/V005596/1, ST/X000672/1, UKRI2492 and ST/C004235/1. 
We thank the GRTL collaboration (\url{www.grtlcollaboration.org}) for their support and code development work. 
This work used the DiRAC Memory Intensive service Cosma8 at Durham University, managed by the Institute for Computational Cosmology on behalf of the STFC DiRAC HPC Facility (\url{www.dirac.ac.uk}). The DiRAC service at Durham was funded by BEIS, UKRI and STFC capital funding, Durham University and STFC operations grants. DiRAC is part of the UKRI Digital Research Infrastructure. 
We acknowledge as well the EuroHPC Joint Undertaking for awarding this project access to the EuroHPC supercomputer LUMI-C, hosted by CSC (Finland) and the LUMI consortium through a EuroHPC
Extreme Scale Access call, under project EHPC-EXT-2025E02-070.

\appendix
\section{Convergence test}
We perform a convergence test using the data from the scalar-damped equal-mass BBH simulations of massive EsGB gravity. The scalar mass and damping constant are chosen as $\mu M=0.1$ and $\kappa_0=0.1M^{-1}$ respectively. A quantitative measure of convergence can be computed from the convergence factor $Q_n=(h^n_{\text{LR}}-h^n_{\text{MR}})/(h^n_{\text{MR}}-h^n_{\text{HR}})$~\cite{Alcubierre}, where LR, MR, HR correspond to the low, medium, high resolutions, $h_i\,\propto\,1/N_i$ relates to the grid spacing used for a given resolution, with $N$ being the number of grid points at the coarsest level and $n$ the order of convergence.

\begin{figure}
    \centering
    \includegraphics[width=1\linewidth]{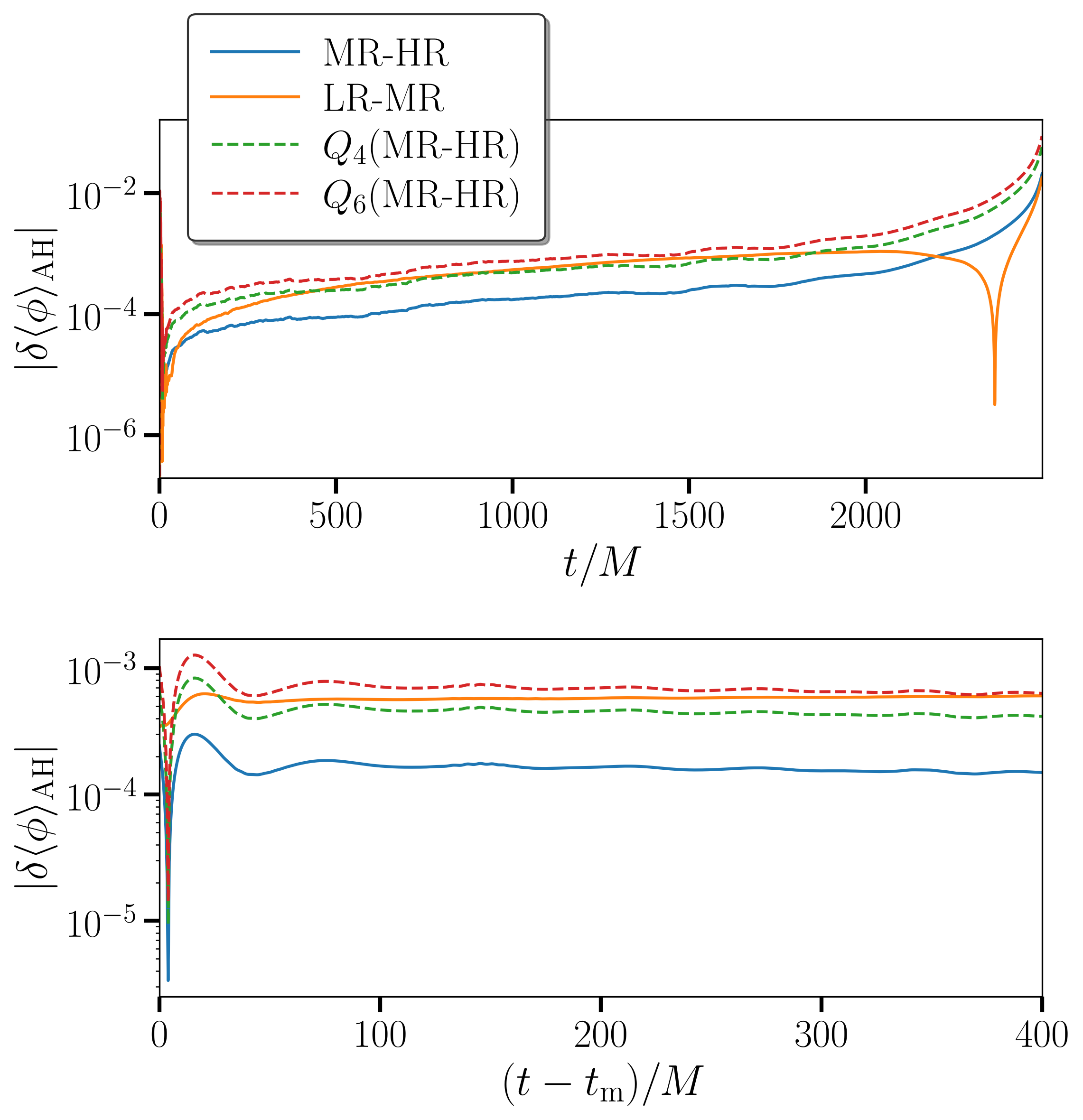}
    \caption{Convergence test performed during the inspiral and post-merger of a BBH merger in massive sGB gravity, using the scalar field extracted at the Apparent Horizon ($\langle\phi\rangle_\mathrm{AH}$) of one of the inspiralling BHs and the remnant BH respectively. We find that the convergence order lies roughly between 4 and 6, which can be expected from the numerical schemes adopted.}
    \label{fig:convergence}
\end{figure}

Using resolutions $N=128, 160$ and 192, we evaluate convergence during the inspiral phase using the scalar field extracted at the Apparent Horizon ($\langle\phi\rangle_\mathrm{AH}$) of one of the BHs (see top panel of Fig.~\ref{fig:convergence}), as well as in the post-merger using the same quantity extracted at the remnant BH (see bottom panel of Fig.~\ref{fig:convergence}), where the simulations are now aligned at merger. At the start of the inspiral, one observes a lower convergence due to the initial scalarisation process of the BHs. Similarly, a lower convergence is found at the late inspiral due to the different merging times for different resolutions. Nevertheless, for the majority of the simulation time, we observe a convergence order between 4 and 6 across the two cases, which is consistent with the fourth-order time evolution and sixth-order spatial discretisation numerical scheme implemented in the code.

\section{Behaviour of the Kretschmann scalar in the strong field regime}\label{sec:Kretsch}

\begin{figure}[H]
    \centering
    \includegraphics[width=1\linewidth]{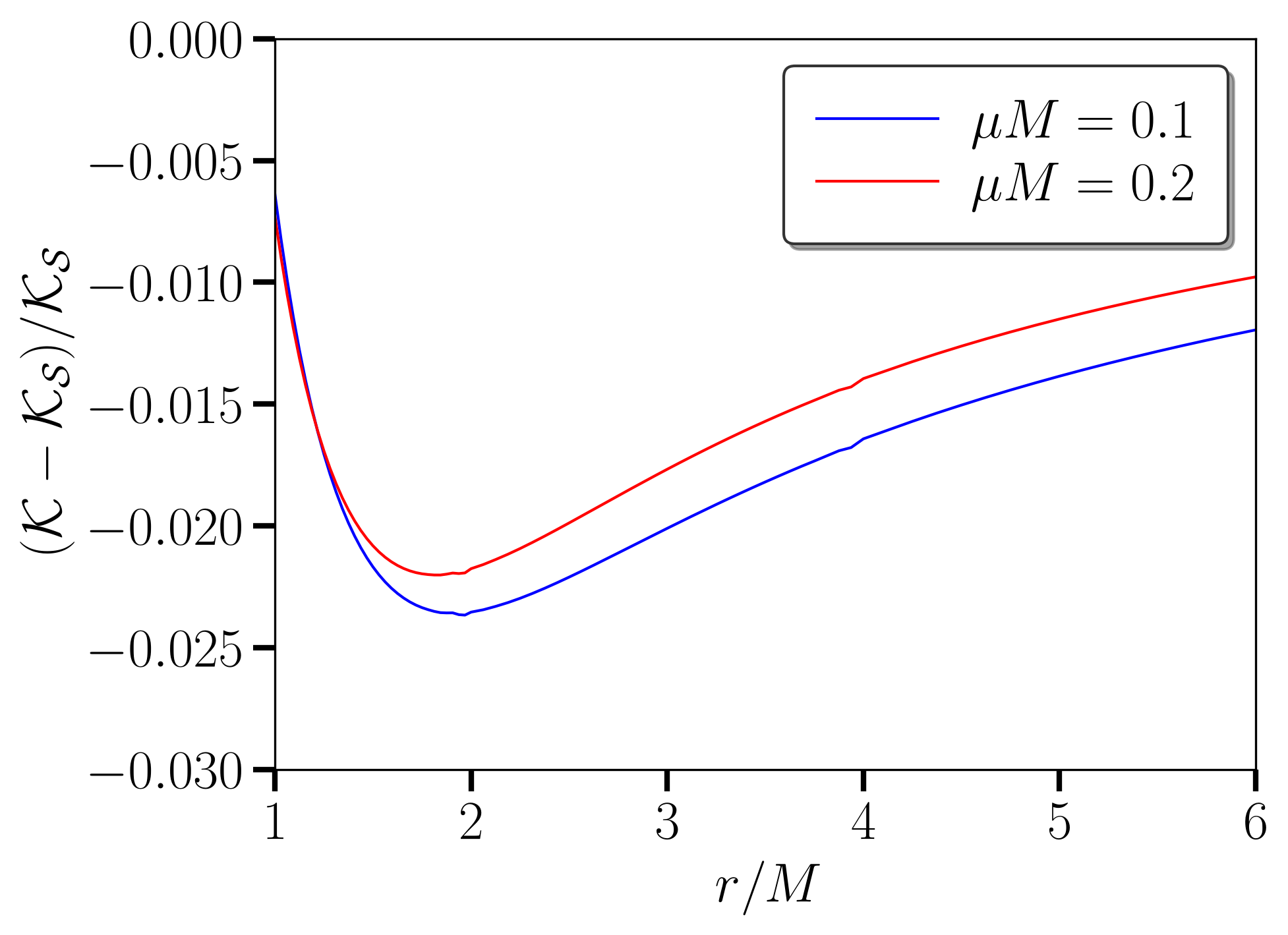}
    \caption{The ratio between $\mathcal{K}-\mathcal{K_S}$ and $\mathcal{K_S}$, where $\mathcal{K}$ is the Kretschmann scalar of a non-rotating BH in massive EsGB gravity and $\mathcal{K_S}$ is that of a Schwarzschild BH in GR, plotted as a function of radial coordinate distance near the apparent horizon. The behaviour of two scalar masses $\mu M=0.1,0.2$ are compared, where we find agreement with the literature that the Kretschmann associated with the heavier scalar lies closer to the Schwarzschild solution. The Kretschmann data used to produce these plots are all extracted at late simulation times ($t=800M$). Note that a numerical solution is also used for the Schwarzschild data (instead of the analytical expression) in order to ensure the comparison is done under the same coordinates.}\label{fig:kretsch}
\end{figure}

We have demonstrated that a massive scalar field can exhibit spurious dynamics if one uses the conventional choice of ID for the scalar ($\phi=\Pi=0$). However, we find that certain geometric variables can still demonstrate reasonable physical behaviour despite the presence of junk dynamics in the scalar field. For example, we can compare the near horizon behaviour of the gauge-invariant Kretschmann scalar $\mathcal{K} \equiv R_{\alpha \nu \rho \sigma}R^{\alpha \nu \rho \sigma}$ in non-rotating BH spacetimes for different choices of scalar mass, analogous to the comparison done in~\cite{10.21468/SciPostPhysCore.7.4.069}. The result is shown in Fig.~\ref{fig:kretsch}, where we find the deviations from the Schwarzschild Kretschmann diminish with an increasing scalar mass, in line with the EFT expectation that a larger scalar mass would suppress corrections to GR. We note that this is also in qualitative agreement with~\cite{10.21468/SciPostPhysCore.7.4.069}, though our different set-ups (e.g. different choice of theory coupling function, coordinate system, etc.) complicate a further quantitative comparison. We believe that one can still obtain physically reasonable results from such comparisons because the distances concerned in this comparison are shorter than the wavelength of the scalar oscillations, making the comparison less sensitive to the spurious scalar dynamics.

\section{Choosing fitting windows}\label{sec:fittingwindows}
Due to the short duration of the ringdown, we consider the following two strategies in choosing our fitting windows for the Fourier-based approaches: i) using a window that starts near a node (i.e. the ``zeros'' in the waveform) at the start of the ringdown and ends near a node towards the end of the ringdown (at $t \sim 120M$ for gravitational waves, and can reach up to $t \sim 250M$ for scalar waves that exhibit a slower decay), or ii) a window starting near a node early in the ringdown but ends at late times ($t \sim 300M$) where the data become noisy. The second class of windows has the advantage over the first class in its ability to produce a smoother spectrum due to its inclusion of a larger number of data points, though at the expense of incorporating late-time noise into the fitting window. In order to use the second class of fitting windows, we must verify that the effect of including noise in our fitting window does not significantly impact the extracted frequencies. This is achieved by comparing the two methods using the GR gravitational wave (2,2) mode ringdown data. For each class of fitting windows, we vary the window size by modifying the starting and/or ending times (given that the window still starts near a node of the waveform) to ensure that the results are stable against fitting window changes. While fitting windows of type (i) found an average peak frequency of $M_{\rm f}\cdot\omega \approx0.522$, windows of type (ii) obtained a similar result of $M_{\rm f}\cdot\omega \approx0.523$. Both results agree well with the frequency extracted from time domain fitting (see table \ref{tab:GW22}), as well as the expected value of $M_{\rm f}\cdot\omega=0.524$ for a spin-2 perturbation of a Kerr BH. This suggests that including noisy data into our fitting window does not significantly impact the extracted peak frequencies, and Fourier-based methods can produce reasonably accurate results for our purposes. We further tested the Fourier-based methods against the time domain fitting results using gravitational wave (2,2) ringdown data from the other EsGB simulations, where we find a strong agreement between the two (see Table \ref{tab:FT_benchmarking}).

For the waveforms we consider in this analysis, we will adopt fitting windows of class (ii) if the ringdown data is too short (e.g. $t_{\rm end}-t_m \leq 150M$), and windows of type (i) if the noise impacts the data at sufficiently late times (e.g. some of the scalar wave modes).

\begin{table}[ht]
\centering
\begin{tabular}{c c c}
\hline
 & $M_{\rm f}\cdot\omega_{1mode}$ & $M_{\rm f}\cdot\omega_{fourier}$ \\
\hline
GR   & 0.525 & 0.523\\
$\mu M=0$   & 0.526 & 0.524 \\
$\mu M=0.1$ & 0.530 & 0.529 \\
$\mu M=0.2$ & 0.536 & 0.535 \\
\hline
\end{tabular}
\caption{Extracted frequencies for the ringdown of (2,2) mode gravitational waves in the different theories considered. The frequencies are extracted from a one-mode fitting of the time domain waveform ($M_{\rm f}\cdot\omega_{1mode}$) and a Fourier-based frequency analysis ($M_{\rm f}\cdot\omega_{fourier}$) respectively, where a strong agreement is found.}
\label{tab:FT_benchmarking}
\end{table}
\bibliography{Bib}
\end{document}